\documentclass[aps,prd,10pt,notitlepage,nofootinbib,superscriptaddress,showkeys,showpacs]{revtex4-1}

\usepackage{amstext,amsmath,amssymb,amsfonts,bbm}
\usepackage[latin1]{inputenc}
\usepackage{epsfig}
\usepackage{hyperref}
\usepackage{amsthm}
\usepackage{subfigure}
\usepackage{color}
\usepackage{multirow}

\usepackage{latexsym}

\usepackage{rotating}

\usepackage{cancel}

\newcommand{\bea}{\begin{eqnarray}}	
\newcommand{\eea}{\end{eqnarray}}
\newcommand{\beq}{\begin{equation}}	
\newcommand{\eeq}{\end{equation}}

\newcommand{\N}{\mathbb{N}}
\newcommand{\C}{\mathbb{C}}
\newcommand{\R}{\mathbb{R}}
\newcommand{\Z}{\mathbb{Z}}

\newcommand{\cG}{{\mathcal G}}
\newcommand{\cV}{{\mathcal V}}
\newcommand{\cL}{{\mathcal L}}

\newcommand{\cF}{{\mathcal F}}
\newcommand{\bG}{{\partial\mathcal G}}

\newcommand{\ren}{\text{ren\,}}

\newcommand{\kin}{{\rm kin\,}}
\newcommand{\inter}{{\rm int\,}}
\newcommand{\ext}{{\rm ext\,}}

\newtheorem{lemma}{Lemma}

\newtheorem{theorem}{Theorem}
\newtheorem{proposition}{Proposition}

\begin{document}

\title{Some classes of renormalizable tensor models}

\author{{\bf Joseph Ben Geloun}}
\email{jbengeloun@perimeterinstitute.ca}
\affiliation{Perimeter Institute for Theoretical Physics, 31 Caroline
St, Waterloo, ON, Canada}
\affiliation{International Chair in Mathematical Physics and Applications,
ICMPA-UNESCO Chair, 072BP50, Cotonou, Rep. of Benin}

\author{{\bf Etera R. Livine}}
\email{etera.livine@ens-lyon.fr}
\affiliation{Laboratoire de Physique, ENS Lyon, CNRS-UMR 5672, 46 All\'ee d'Italie, Lyon 69007, France}
\affiliation{Perimeter Institute for Theoretical Physics, 31 Caroline
St, Waterloo, ON, Canada}

\date{\small\today}

\begin{abstract}
\noindent
We identify new families of renormalizable of tensor models
from anterior renormalizable tensor models via a
mapping capable of reducing or increasing the rank of the
theory without having an effect on the renormalizability
property. Mainly, a version of the rank 3 tensor model as defined in
[arXiv:1201.0176 [hep-th]], the Grosse-Wulkenhaar model in 4D and 2D generate three different classes of renormalizable models. 
The proof of the renormalizability is fully performed for the first reduced
model. The same procedure can be applied for the remaining cases. Interestingly, we find that, due to the peculiar behavior of anisotropic wave function renormalizations, the rank 3 tensor model reduced to a matrix model generates a simple super-renormalizable vector model. 

\medskip
\noindent Pacs numbers: 11.10.Gh, 04.60.-m, 02.10.Ox
pi-qg-285 and ICMPA/MPA/2012/12 \\
\today 

\end{abstract}

\maketitle

\tableofcontents

\section{Introduction}

Tensorial Group Field Theory (TGFT) \cite{oriti,Rivasseau:2011xg,Rivasseau:2011hm,Rivasseau:2012yp}
is a recently built quantum field theoretical framework pertaining to the ``discrete to continuum'' scenario for quantum gravity. In such an  instance, fields are tensors
of rank $D$ labeled by abstract group representations which are viewed as $D$ simplexes and the interactions are of the form of $D+1$ simplexes.  
According to quantum field theory rules, tensor fields of rank $D$ or $D$ simplexes are glued along their $D-1$ simplexes and interact in $D+1$ simplexes. The simplicial approach for discussing quantum gravity is in fact well known and has led to the famous study of matrix models in lower dimensional statistical mechanics \cite{Di Francesco:1993nw,nwambj3dqg}. TGFT should be regarded as the higher rank extension of random matrix models with a bonus: the genuine feature to be renormalizable. 

Within the above TGFT framework, new classes of renormalizable models involving tensor fields have been highlighted  \cite{BenGeloun:2011rc,BenGeloun:2012pu,BenGeloun:2012yk,
Samary:2012bw,Carrozza:2012uv,Carrozza:2013wda}. 
The renormalization procedure for these models involves an extended notion  of multiscale analysis \cite{Rivasseau:1991ub} which steers a new kind power counting theorem and locality principle. It is noteworthy that all interactions are nonlocal (precisely, they happen in a region of the abstract group manifold or position space) and build around
Gurau's $1/N$ expansion \cite{Gur0,Gur4,Gur5,Gur6,Bonzom:2011zz,Gurau:2011kk,Dartois:2013he,Gurau:2013cbh} for higher rank colored theories \cite{color,Gurau:2011xp}. 
Remarkably, renormalizable TGFTs shed as well more light on anterior results in renormalization of quantum matrix models on noncommutative spaces \cite{Rivasseau:2007ab,Grosse:2004yu,Grosse:2003nw}.
Indeed, at the perturbative level, Feynman graphs of TGFT models are generated by vertices and propagators spanned by stranded graphs representing higher rank extension of ribbon vertices and propagators as used in the matrix formulation of the Grosse-Wulkenhaar (GW) model \cite{Grosse:2004yu,Grosse:2003nw}. Power counting theorems and locality
principle in TGFT extend several similar notions found in the matrix case.

At quantum gravity energy scale, some axioms or principles of ordinary quantum theory should be, if not drastically revised, at least profoundly rethought. Several theoretical frameworks address the fact that locality should
no longer hold in that regime (string theory, noncommutative geometry,  etc...). Let us focus on the particular forms of nonlocality as appear in TGFTs. 
In a broader sense, the models cited above belong to the class of models endowed with nonlocal interactions.
Field arguments in the interaction term might be paired in
many possible ways. Specific forms have to be physically motivated and tractable. For instance, in noncommutative field theory
induced by noncommutative Moyal field algebra, four fields
interact in a region (a parallelogram) the area of which is
the Planck length square \cite{Rivasseau:2007ab}. The recent tensor models in \cite{BenGeloun:2012pu, BenGeloun:2011rc} possess
interactions of the form of 3 and 4-simplex which generate, through a path integral formalism, simplicial pseudo-manifolds in 3D and 4D.

In a nonlocal field theory, it could happen that
the interaction is of a definite form which
can be called ``partially cyclic'', namely, in the interaction pairing, a tensor field only share (at least two) arguments
with at most two other fields (a ``totally cyclic'' or simply ``cyclic''
interaction would be an interaction having this property valid for all
fields which define it). Consider some tensors
$T^i$ and an interaction defined by contractions of indices of the $T^i$'s such that
\beq
I=\sum_{[J],[K],\dots} T^1_{\dots, [J],\dots}T^2_{\dots, [J],[K],\dots}
T^3_{\dots, [K],\dots}\dots\,,
\eeq 
where $[J]$ and $[K]$ are block
indices or arguments. Consider now a similar but simpler interaction
where indices $i$ and $k$ replace block indices $[J]$ and $[K]$ such that
\beq
\tilde I=\sum_{j,k,\dots}T^1_{\dots,j,\dots}T^2_{\dots,j,k,\dots} T^3_{\dots,k,\dots}\dots.
\eeq
It is natural to ask: how starting from $I$, one generates $\tilde I$ and what is the main feature of the reduced model  described by $\tilde I$? A quantum field theory being not only defined by interactions
what implications has such a reduction on the dynamics?
These questions might be very intricate
and, more to the point, even more complex if one would like to preserve some nice properties such as renormalizability or symmetry  aspects of the initial theory.
If all features of the model described by $I$ are fully represented 
in $\tilde I$, then $\tilde I$ could provide a much simpler model than $I$. 

These questions could find also an importance in the  double scaling limit analysis of tensor models \cite{Gurau:2011sk}. In this latter work, the author projects the rank $D$ tensor random variables $T_{a_1,a_2,\dots,a_D}$ onto a simplified rank 2
tensor $\tilde T_{a_1,\vec a_2}$ so that particular cyclic tensor interactions can be 
merely seen as matrix trace invariants. This has led to the discovery of new 
multi-critical points and a novel double scaling limit for matrix models. 

Several points about TGFT have yet to be addressed. TGFT
might generate a wealth models with several possible interactions and kinetic terms. 
We definitely need a guidance towards true physical models which should incorporate
the geometry seed of a theory of General Relativity. Including renormalizability in the game might be one selection criterion. Under the Renormalization Group (RG) flow, only microscopic physically robust models with long-lived logarithmic flow couplings would resist to the several layer scales and would finally give  macroscopic observable effects. This is indeed what a quantum theory for gravity would demand 
and this is indeed why our first intention is to preserve the renormalizability feature 
of any TGFT models generated. In particular, if the reduction $I \to \tilde I$ respects that feature then it could provide an important simplification
worth to be investigated. 

In this paper, we show that, at least three particular
classes of tensor models equipped with cyclic interactions
can be projected back to  reduced rank models and, reciprocally,
any tensor model of this kind can be extended to a higher rank tensor model with a cyclic interaction in a sector.
During the process, we are able to identify new classes of renormalizable models (perturbatively and at all orders). If the projection and extension
of these models can be somehow understood
either by dividing or by multiplying the number of indices,
what we actually show is that this mapping
preserves the renormalizability of the initial model.
The proof is fully established for a reduced rank 3 tensor model
issued from \cite{BenGeloun:2012pu}. The model considered is an independent not identically distributed random matrix model. In other words, it is anisotropic in the sense that its strands are not equally weighted from the point of view of the measure (the GW model with a magnetic field \cite{Geloun:2008zk} is likewise but not totally similar).  
The renormalization procedure for tensor models is always a delicate issue.
However, we show here that the full renormalization program (from the multi-scale analysis to the renormalization of divergences) applies for this peculiar model. 
The proof also applies to the GW model in 2D \cite{Grosse:2003nw} and 4D \cite{Grosse:2004yu},
and, from these, the universal feature of our formalism can be easily inferred. 
Theorem \ref{theo1} and Theorem \ref{theo3} are our main results
and they mainly allow us to identify three different families of renormalizable 
tensor models. Interestingly, we find that there exists a continuum of
perturbatively renormalizable theories linking the three classes. 

We would like to stress also the fact that the projection-extension mechanism 
can be used to reduce a piece of the rank 4 model of \cite{BenGeloun:2011rc}  but cannot reduce the entire interaction to a matrix interaction. 
Thus, there exist actually renormalizable tensor models which are not partially cyclic. 
Another point that has been also not entirely covered in this paper
is whether or not the procedure generates models stable under
the RG flow. In some situations, 
it may actually happen that the initial model flows towards 
a different model hence it is not stable. However, in all situations carried
out below, the choice of coupling constants of the model is made
in such a way that this odd feature is simply avoided. 

The plan of this paper is the following:
In the next section, as a complete test of the above ideas,
we carry out the full renormalization program for a new matrix model built
from a renormalizable rank 3 renormalizable model  \cite{BenGeloun:2012yk}. 
Section \ref{sect:nclass} is devoted to the identification of
new families of renormalizable tensor models issued from GW models
in 4D and 2D by applying the program of Section \ref{sect:rank3}
to these models. 
We infer the existence of three different families of renormalizable
models (rank 3 tensor, GW 4D and GW 2D models) having three different
renormalizable matrix models as roots. 
An appendix discussing the one-loop $\beta$-function or UV behavior of some of the models treated closes the paper.

\section{A just normalizable matrix analogue of a rank three tensor model}
\label{sect:rank3}

\subsection{Rank 3 tensor model and its matrix reduction}

Let us recall the main features of the the model defined in \cite{BenGeloun:2012pu} henceforth called $T3$ model. The following results and transformations  will find consistent analogues for other types of tensor models discussed in the remaining sections.

Consider complex fields $\varphi : U(1)^3 \to \C$ which can be
equivalently described after Fourier mode decomposition as tensors 
\beq
\varphi(g_1,g_2,g_3)= \sum_{p_i} \varphi_{[p]} \,e^{ip_1\theta_1}
e^{ip_2\theta_2}e^{ip_3\theta_3}\,,
\qquad g_k=e^{ip_k\theta_k}\,, \qquad \theta_k\in[0,2\pi)\,,
\eeq
where $[p]= (p_1,p_2,p_3)$, $p_i \in \Z$. 
It is possible to restrict the discussion for positive mode fields, i.e.
we assume that tensor fields satisfying the symmetry
\beq
\varphi_{p_1, p_2,p_3} = \varphi_{-p_1,p_2,p_3} =
\varphi_{p_1,- p_2,p_3} =  \varphi_{p_1,p_2,-p_3} \,.
\label{modesym}
\eeq
Thus, we will consider only fields such that $p_i \in \N$, namely
fields can  be regarded as living in $(U(1)/\Z_2)^3$.
Note such a restriction is made by sake of simplicity.
Hence, such a prescription will have no
consequence on the subsequent analysis (this point will be emphasized later on).

\begin{figure}[ht]
 \centering
      \centering
\includegraphics[angle=0, width=3cm, height=0.5cm]{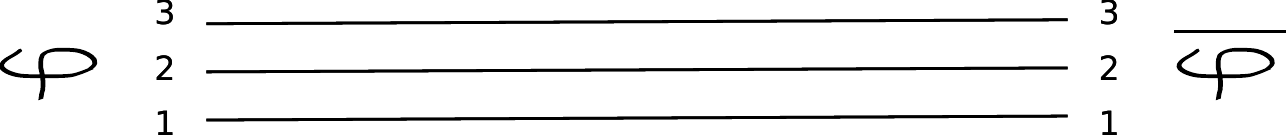}
\vspace{0.5cm}
\includegraphics[angle=0, width=13cm, height=3.5cm]{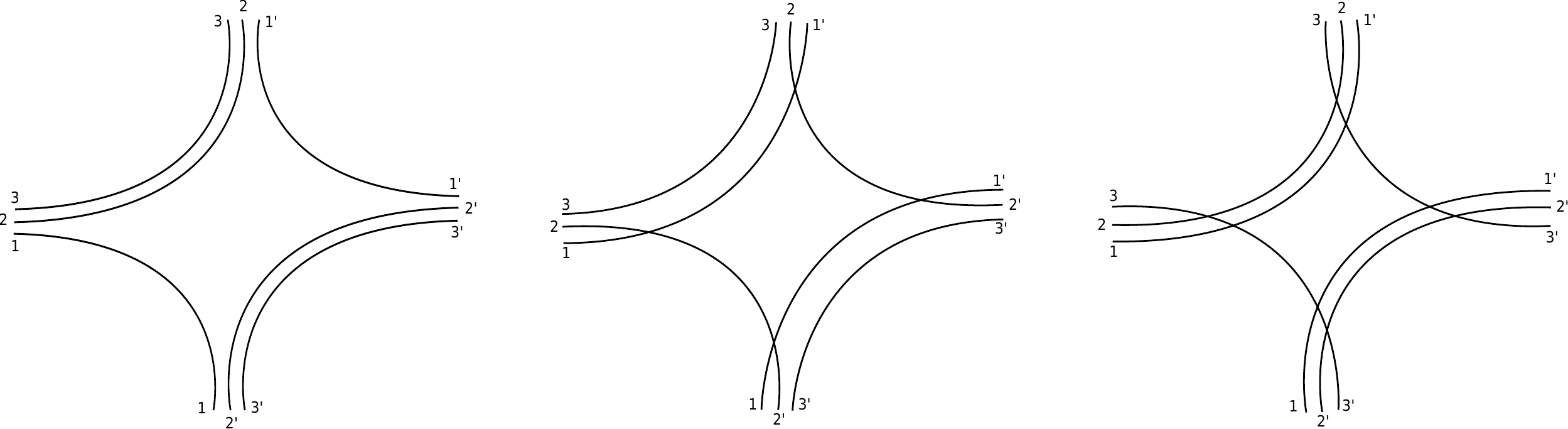}
\caption{ {\small Propagator and vertices of the type $\varphi^{4}$ of
the rank 3 tensor model. }}
\label{fig:vertexinit}
\end{figure}

Using now these tensor components such that
$\varphi_{p_1,p_2,p_3} \in \C$, $p_i \in \N$, the $T3$ model possesses
a kinetic term given by
\beq
S^{\kin} =\sum_{p_j} \bar{\varphi}_{p_1,p_2,p_3}\Big(
\sum_{s=1}^3 a_sp_s + \mu \Big)\varphi_{p_1,p_2,p_3},
\label{eq:action0}
\eeq
with some mass $\mu$, $a_s$ are wave-function couplings
associated with the theory propagator 
\beq
\widehat C([p_i];[\tilde p_i])=
\prod_{i=1}^3\delta_{p_i, \tilde p_i}/\big(\sum_{s=1}^3a_sp_s + \mu\big)\,.
\eeq
Hence the Gaussian measure $d\mu_C[\varphi]$ of the model 
has a covariance 
$C = 1/(\sum_{s=1}^3a_sp_s + \mu)$. 
The interaction of the $T3$ model is of $\varphi^4$-type given by  
\beq
S^{\inter}=\sum_{p_j}\varphi_{p_1,p_2,p_3}\,\bar{\varphi}_{p_{1'},p_2,p_3}\,
\varphi_{p_{1'},p_{2'},p_{3'}}\,\bar{\varphi}_{p_{1},p_{2'},p_{3'}} +\;
\text{permutations}\,, 
\label{vertex4}
\eeq
where ``permutations'' refers to other terms induced by 
color symmetry on strand indices. 
Propagator and vertices of the model are pictured in Fig.\ref{fig:vertexinit}.

On of the main theorem proved in \cite{BenGeloun:2012pu} consists
in the following statement: Introducing a UV cut-off $\Lambda$ on the propagator $C \to C^\Lambda$, the multi-scale analysis of graph amplitudes \cite{Rivasseau:1991ub} proves that the action 
\beq
S^\Lambda = \lambda^\Lambda S^{\inter} + 
CT^{\Lambda}_{2;1} S_{2;1} +\sum_{s=1,\dots,3} CT^{\Lambda}_{s;22}
S_{s;2;2} \,, 
\eeq
where $CT^{\Lambda}_{2;1} S_{2;1}$ is a mass counter-term and
$CT^{\Lambda}_{s;22} S_{s;2;2}$, $s=1,2,3$, are wave-function
counter-terms, and the related a partition function
\bea
\mathcal{Z} = \int d\mu_{C}^{\Lambda}[\varphi] e^{-S^\Lambda}
\eea
define a model which is renormalizable at all orders of perturbation theory.

We introduce now the ``anisotropic'' model defined by the 
interaction
\beq
S_{T3}^{\inter}=\sum_{p_j}\varphi_{p_1,p_2,p_3}\,\bar{\varphi}_{p_{1'},p_2,p_3}\,
\varphi_{p_{1'},p_{2'},p_{3'}}\,\bar{\varphi}_{p_{1},p_{2'},p_{3'}} \,, 
\label{vertex4}
\eeq
and keeping still the same kinetic term given by \eqref{eq:action0}. 
Thus, we have explicitly broken the strand
symmetry by choosing such an interaction. This breaking enforces 
a particular choice of wave function couplings (see discussion of Section 5.3 in \cite{BenGeloun:2012pu}) affording a proper notion 
of wave function renormalization. 
 The choice of $a_1\neq a_2 =a_3$ is therefore
of great significance because it is only under these conditions that
the model defined by $S^{\kin}$ and $\lambda S^{\inter}_{T3}$ turns out to be just renormalizable \cite{BenGeloun:2012pu}. 
In other words, other choices might lead to instability under the RG flow.  

Henceforth, we will restrict our analysis 
to the model with the unique interaction \eqref{vertex4}.
This is nothing but the first vertex depicted in Fig. \ref{fig:vertexinit}.

A striking feature of the interaction \eqref{vertex4}
is that it maps to a pure  matrix interaction
using any bijection $\tilde\sigma:\N^2 \to \N$.
 Indeed, consider
the following field redefinition:
\beq
\varphi_{p_1,p_2,p_3} \;\mapsto\; \phi_{p, n}\,,
\qquad p_1 = p\,, \quad \tilde\sigma(p_2,p_3)=n
\eeq
to which, given $\tilde\sigma(p,q)=n$ and its inverse noted as $\tilde\sigma^{-1}_1(n)=p$
and $\tilde\sigma^{-1}_2(n)=q$, corresponds the following transformed actions  (from now on, $a_1=a$, $a_2=a_3=b$),
\bea
S_{T3}^{\kin} &=& \sum_{p,n} \bar{\phi}_{p,n}\Big(ap +b(\tilde\sigma^{-1}_1(n)+\tilde\sigma^{-1}_2(n))+\mu\Big)\phi_{p,n} \,,
\label{nkin}\\
S_{T3}^{\inter}&=&
\sum_{p_1,p_2,n_1,n_2}\phi_{p_1,n_1}\,\bar{\phi}_{p_{2},n_1}\,
\phi_{p_{2},n_2}\,\bar{\phi}_{p_{1},n_2}\,.
\label{ninter}
\eea
We reduce the $(p_2,p_3)$-momentum sector according to the fact that the interaction is cyclic with respect to this couple of indices.

\begin{figure}[ht]
 \centering
        \centering
\vspace{-0.5cm}
\includegraphics[angle=0, width=3cm, height=0.5cm]{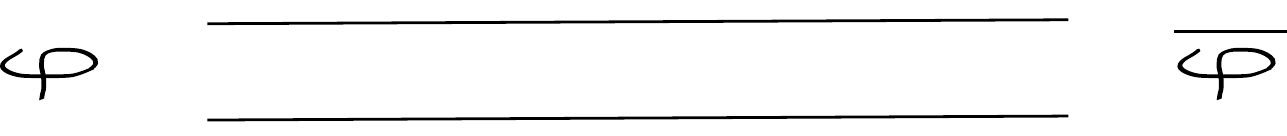}
\hspace{0.5cm}
\includegraphics[angle=0, width=4cm, height=2.5cm]{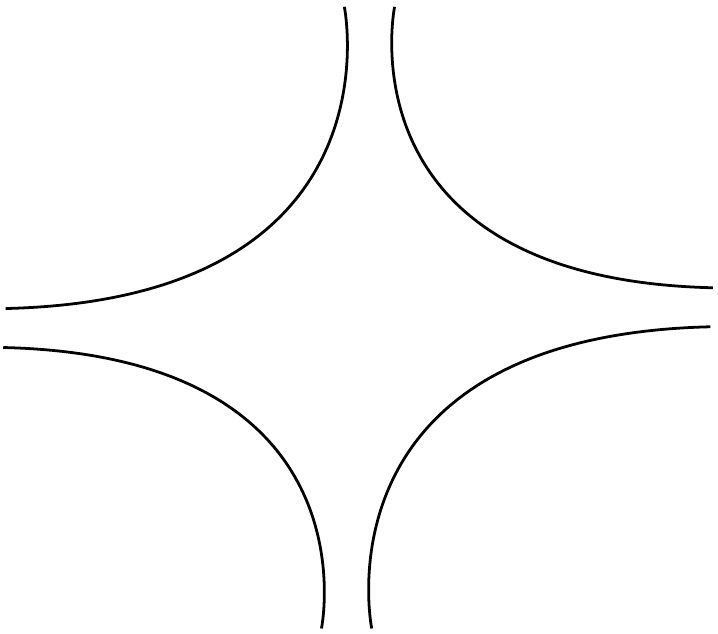}
\caption{ {\small Propagator and vertex in the reduced matrix theory. }}
\label{fig:vertex2d}
\end{figure}

Being unique and of the matrix kind, the interaction \eqref{ninter}   cannot clearly generate under the RG flow any other coupling than itself.
It can be represented in the ordinary form of Fig.\ref{fig:vertex2d}.

\begin{figure}[h]
\begin{minipage}[t]{.8\textwidth}
\begin{center}
\includegraphics[height=30mm]{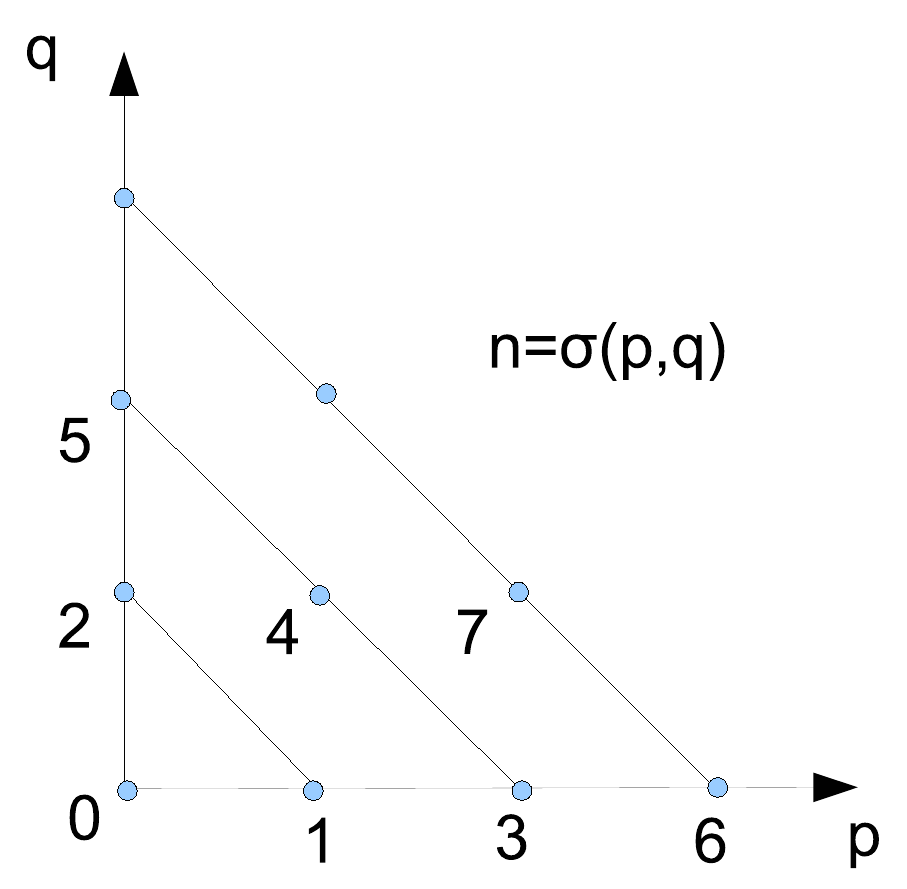}
\hspace{5mm}
\includegraphics[height=30mm]{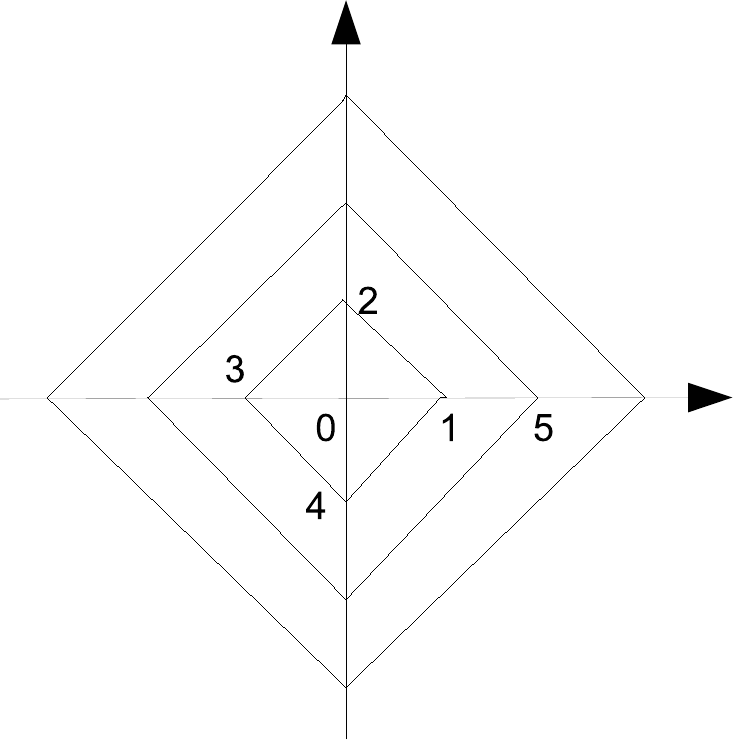}
\caption{ {\small On the left, the bijection $\sigma$ labeling all pairs of integers and mapping $\N^2$ onto $\N$, and on the right its generalization to a bijective map from $\Z^2$ onto $\N$. }\label{sigmamap}}
\end{center}
\end{minipage}
\end{figure}

We now investigate the implications induced by the reduction procedure on the kinetic term. We introduce
\beq
N_n = \tilde\sigma^{-1}_1(n)+\tilde\sigma^{-1}_2(n)\,.
\eeq
In particular, the standard choice $\sigma$ for $\tilde \sigma$ is defined by ordering the pairs of integers $(p,q)$ in $\N^2$ along the diagonal (at constant $p+q$) and numbering them from bottom to top, as illustrated in Fig.\ref{sigmamap}.

Explicitly, this map reads:
\beq
\sigma(p,q) = \frac{1}{2}(p+q)(p+q+1) + q\,,\qquad 
\sigma(0,0) = 0
\eeq
and its inverse can be characterized as
\beq
\sigma^{-1}(n) = (p,q)\in \N^2\,,  \quad
N = p+q \,,   \quad N(N+1) \leq 2n \leq N(N+3)\,,
\quad q = 2n - N(N+1) \,, \quad  p = N - q \,,
\label{sigm}
\eeq
where  the inequality uniquely determines
$N=N_n\in \N$. We denote $\sigma^{-1}_1(n)= p$ and
$\sigma^{-1}_2(n)= q$. It can be easily shown that the
following (optimal) bounds holds:
\bea
&&
\frac{\sqrt{9+ 8n}- 3}{2}\leq N \leq  \frac{\sqrt{1+ 8n} - 1 }{2}\,, \crcr
&&
\frac{(\sqrt{9+ 8n}- 3)^2}{8}\leq \frac{N^2}{2}\leq  p^2 + q^2 \leq
\Big[ (p+q)^2 = N^2 \Big]\leq  \frac{(\sqrt{1+ 8n} - 1)^2 }{4} \leq 2n\,.
\label{boundN}
\eea
Hence,  for large $n$ (such a condition will find a motivation later as 
the UV limit where field modes proliferate),
\bea
\sqrt{2n}- \frac32\leq  \sqrt{2n}\Big[ \sqrt{\frac{9}{8n}+1} - \frac{3}{\sqrt{8n}}\Big] \leq N \leq  \frac{\sqrt{1+ 8n} - 1 }{2} \leq \sqrt{2n}\,,
\eea
the approximation $N \sim \sqrt{2n}$ is correct. We could of course introduce another choice of bijection between $\N^2$ and $\N$, but the map $\sigma$ is, in some sense, the most compact choice for which we have a natural simple estimate of $N=p+q$  in terms of $n=\sigma(p,q)$. For a more general map, we would reshuffle the labeling of points $(p,q)\in\N^2$ by the integer $n$, which would lead to a more random behavior of $N$ that would wildly fluctuate away from $\sqrt{2n}$.

Let us point out that we can easily adapt our analysis to $\Z^2$ and introduce a generalized map $\sigma:\Z^2\rightarrow \N$ as shown in Fig.\ref{sigmamap}, labeling points $(p,q)\in\Z^2$ along the ``circles" of constant $N=|p|+|q|$. This would lead to a similar behavior of $N$ scaling proportionality to $\sqrt{n}$.
 This also sustains the fact that we could have let all field
modes $p_i\in \Z$ (without assuming any symmetry \eqref{modesym}) and still the analysis will be valid.

What boils down in the above algebra is that we can introduce the following kinetic term 
\bea
S^{\kin}_{\sigma T3} = \sum_{p,n} \bar{\phi}_{p,n}\Big(ap +b\sqrt{n}+\mu\Big)\phi_{p,n}
\eea
associated with a new propagator (which draws as in Figure \ref{fig:vertex2d})
\beq
\widehat C([p,n];[\tilde p,\tilde n])=
\delta_{p, \tilde p}\delta_{n, \tilde n}/\big(ap + b\sqrt{n} + \mu\big)\,.
\label{eq:propn}
\eeq
Next we introduce a UV cut-off $\Lambda$ on the new propagator
 \eqref{eq:propn} and $CT^\Lambda _{2;0} S_{2;0}$ a mass counter-term 
and $CT^\Lambda _{2;s} S_{2;s}$, $s=1,2$,  two wave function counter-terms defined as 
\beq
S_{2;0} = \sum_{p,n} \phi_{p,n} \phi_{p,n}\,, \quad
S_{2;1}=\sum_{p,n} \phi_{p,n} \,p\, \phi_{p,n}\,, \quad 
S_{2;2}= \sum_{p,n} \phi_{p,n}\, \sqrt{n} \,\phi_{p,n}\,.
\label{wfct}
\eeq 
Our remaining task is to prove that, after the tensor-to-matrix reduction procedure followed by the approximation of $N_n \sim \sqrt{2n}$,
the following statement holds: 

\begin{theorem}[Renormalizability of $\sigma T3$]
\label{theo1}
The $\sigma T3$ model defined by
\beq
S^\Lambda = \lambda^\Lambda S^{\inter}_{T3}
 + \sum_{s=0,1,2} CT^\Lambda _{2;s} S_{2;s}
\eeq
with Gaussian measure $d\mu_C^\Lambda(\phi)$ 
with covariance associated with the kinetic term $S^{\kin}_{\sigma T3}$,
 is renormalizable at all orders of perturbation theory. 
\end{theorem}

The proof of this statement will follow a multi-scale analysis
leading to a power counting theorem and renormalization
procedure for divergent terms in the way of \cite{Rivasseau:1991ub}.

\subsection{Multiscale analysis}
\label{subsect:multi}

\noindent{\bf Propagator bound.} Using Schwinger's kernel, we rewrite the propagator \eqref{eq:propn}  and its slice decomposition \cite{Rivasseau:1991ub} as
\bea
&&
C(p,n) =\frac{1}{ a p +b\sqrt{n}+\mu} =
\int_0^\infty d\alpha  \, e^{-\alpha(ap + b\sqrt{n} + \mu)} \,,\crcr
&&
C_i(p,n) =
\int_{M^{-i}}^{M^{-i+1}} d\alpha\,   e^{-\alpha(ap +b\sqrt{n} + \mu)}\,,
\qquad
C_0(p,n) =
\int_{1}^{\infty} d\alpha\,   e^{-\alpha(ap +  b\sqrt{n}  + \mu)}\,. 
\eea
for some large constant $M \in \N$. One has $C = \sum_{i=0}^\infty C_i.$
We get the following bounds on the sliced propagators:
\beq
C_i(p,n)\leq K M^{-i} e^{- M^{-i}(ap+ b\sqrt{n}  +\mu)}\,,
\qquad
C_0(p,n)  \leq K  e^{- (ap+ b\sqrt{n} +\mu)}\,, 
\label{ci}
\eeq
for  $K\geq 0$. 
Hence, a slice $i\gg 1$ probe high momenta either of
order $M^{i}$ or of order $M^{2i}$. A UV cut-off can be introduced such that the cut-offed propagator is $C^\Lambda = \sum_{i=0}^\Lambda C_i$ and the UV limit is obtained
by taking $\Lambda \to \infty$. It is common to 
refer $C_0$ and $C_\Lambda$ to as the IR and UV propagator
slice, respectively. The supscript $\Lambda$ will be dropped
in the following. 

\medskip
\noindent{\bf Optimal bound on an amplitude.}
Given a connected graph $\cG$ with set of vertices 
$\cV$, $V = |\cV|$, set of lines $\cL$, $L= |\cL|$, 
for evaluating the optimal bound on the bare amplitude associated with such a graph, we proceed in the usual way \cite{Rivasseau:1991ub} but we must take into account the effect introduced by the new propagator. First write the bare amplitude of the graph as
\bea
A_\cG = \sum_{\mu} f_{\mu}(\lambda, CT) A_{\cG,\mu}\,,
\qquad
A_{\cG,\mu}  = \sum_{p_{v,s}} \Big[\prod_{\ell \in \cL} C_{i_\ell(\mu)}(\{p_{v(\ell),s}\};\{p_{v'(\ell),s}\})\Big]\prod_{v \in \cV; s.s'} \delta_{p_{v,s},p_{v,s'}}\,,
\eea
where $\mu=\mu(i_1, i_2,\dots, i_q)$ is called 
momentum assignment and gives to each propagator 
of each internal line $\ell$ a scale $i_\ell \in [0,\Lambda]$; the sum over 
$\mu$ is performed on all assignments and can only be done after renormalization in the way of \cite{Rivasseau:1991ub}. The function $f_\mu$ contains product of coupling constants as well as symmetry factor of the graph. Given a line $\ell$, its propagator momenta $p_{v(\ell);s}$ relate  a vertex $v(\ell)$ to another vertex $v'(\ell)$ and possess also a strand index $s=1,2$. The vertex operator is simply a collection of delta functions identifying entering and exiting momenta. If we do have external lines hooked on the graph $\cG$, we could fix all external line indices to $i_{\ext}=-1$. 

The next stage is to perform in a optimal way the sum on $p_{v,s}$
thereby getting an optimal bound on $A_{\cG,\mu}$ which
is the quantity of interest.  
This optimal sum can be done by introducing the so-called quasi-local subgraphs,
key ingredients for the multiscale analysis \cite{Rivasseau:1991ub}. Given $\mu$ and a scale $i$, we consider the complete list of the connected components $G^{k}_{i}$, $k = 1,2,\dots, k(i)$ of the subgraph $\cG_i$ made of all lines in $\cG$ with the scale attribution $j \leq i$ in $\mu$, with $\cG_0 = \cG$. 
The set of $\{G^k_i\}_{k,i}$ is partially ordered by inclusion. 
The Gallavotti-Nicol\`o tree \cite{galla} is an abstract tree made with nodes the $G^k_i$'s associated with that partial order such that there is a link between two nodes if and only if one is included in the other. Such
a tree has obviously a root $\cG_0=\cG$. We refer the reader 
to Figure 3 in \cite{BenGeloun:2012pu} for a complete illustration 
of this tree in the T3 model from which the similar notion in the present
reduced framework should be clear. The key point
is to choose a spanning tree of lines in the graph $\cG$ and to perform the sum associated with momenta of these lines in such a way ``to be compatible'' with the Gallavotti-Nicol\`o tree. This compatibility condition entails an optimal bound which must be specified. 

The vertex operator and the propagator 
contain both a bunch of delta functions which make that the amplitude 
factorizes along closed and open strands that are called faces. 
Thus the set  $\cF$ of faces divides in $\cF_{\inter}$ set of internal or closed faces with cardinal $F_{\inter}= |\cF_{\inter}|$, and $\cF_{\ext}$ set of external or open faces with cardinal $F_{\ext}= |\cF_{\ext}|$. Moreover, since the momentum associated with such faces can be $p$ or $\sqrt{n}$,
we introduce another discrepancy between the faces: those indexed by a momentum $p_f$ and belonging to the set $\cF^-$ and  those coined by $\sqrt{n_f}$ and belonging to the set $\cF^+$. Consequently, we can also 
have other types of subsets given by 
$\cF^\pm_{\bullet} = \cF^{\pm}\cap \cF_{\bullet}$, $\bullet = \inter,\ext$. 
 We write $|\cF^\pm_{\inter}|= F^\pm_{\inter}$ and $|\cF^\pm_{\ext}|= F^\pm_{\ext}$. 

Then, using the sliced propagator bound \eqref{ci}, after summing over
all delta functions, one comes to   
\bea
|A_\cG| &\leq&  K^{L}  \Big[\prod_{\ell \in \cL} M^{-i_\ell}e^{-M^{-i_\ell} \mu}
\Big]
\sum_{n_f, p_f}  \Big[\prod_{f \in \cF^+} \prod_{\ell \in f} e^{- M^{-i_\ell}b\sqrt{n_f}}\Big]\Big[
\prod_{f \in \cF^-} \prod_{\ell \in f} e^{- M^{-i_\ell}ap_f}
\Big]\crcr
&\leq& K'^{L} \Big[\prod_{\ell \in \cL} M^{-i_\ell}\Big]
\sum_{n_f, p_f}  \Big[\prod_{f \in \cF^+}  e^{- [\sum_{\ell \in f}M^{-i_\ell}]b\sqrt{n_f}}\Big]\Big[
\prod_{f \in \cF^-}  e^{- [\sum_{\ell \in f}M^{-i_\ell}]ap_f}
\Big]
\eea
where the set $\{\ell \in f\}$ denotes the set of strand lines involved
in the face $f$, and $K'$ some constant. We must consider several cases

\begin{enumerate}
\item[(i)] If $f \in \cF^+_{\inter}$, then the face amplitude
is of the form $\sum_{n_f} e^{- [\sum_{\ell \in f}M^{-i_\ell}]b\sqrt{n_f}}$
the sum on $n_f$ can be optimized by choosing $i_f=\min_{\ell \in f} i_\ell$ and 
\beq
\sum_{n_f} e^{-\delta M^{-i} b\sqrt{n_f}} = \delta'M^{2i} + O(M^{i})\,, 
\qquad \delta= |\{\ell \in f\}|\,, \qquad \delta' = \frac{2}{\delta^2 b^2}\,. 
\label{sum1}
\eeq
\item[(ii)] If $f \in \cF^-_{\inter}$, the face amplitude
becomes $\sum_{p_f} e^{- [\sum_{\ell \in f}M^{-i_\ell}]a p_f}$ and it  is optimal by choosing $i_f=\min_{\ell \in f} i_\ell$ such that
\beq
\sum_{p_f} e^{-\delta M^{-i} ap_f} = \delta'M^{i} + O(M^{i})\,, 
\qquad \delta= |\{\ell \in f\}|\,, \qquad \delta' = \frac{1}{\delta a}\,. 
\label{sum2} 
\eeq
\item[(iii)] Assume now that $f \in \cF_{\ext}$, all intermediate momenta can be summed and yield $O(1)$. 
\end{enumerate}
We therefore obtain
\bea
|A_\cG| &\leq& K'^{L} \Big[\prod_{\ell \in \cL} M^{-i_\ell}\Big]
\sum_{n_f, p_f}  \Big[\prod_{f \in \cF_{\inter}^+}  e^{- [\sum_{\ell \in f}M^{-i_\ell}]b\sqrt{n_f}}\Big]\Big[
\prod_{f \in \cF_{\inter}^-}  e^{- [\sum_{\ell \in f}M^{-i_\ell}]ap_f}
\Big]\crcr
&\leq& K'^{L} K''^{F_{\inter}}\Big[\prod_{\ell \in \cL} M^{-i_\ell}\Big]
 \Big[\prod_{f \in \cF_{\inter}^+}  M^{2i_f} \Big]\Big[
\prod_{f \in \cF_{\inter}^-}  M^{i_f} \Big]\,.
\eea
where $K''$ is some constant. The above amplitude rewrites
using the $G^k_i$'s as
\bea
|A_\cG| &\leq& K'^{L} K''^{F_{\inter}}\Big[\prod_{\ell \in \cL}\prod_{i=1}^{i_\ell} M^{-1}\Big]
 \Big[\prod_{f \in \cF_{\inter}^+} \prod_{i=1}^{i_f}  M^{2} \Big]\Big[
\prod_{f \in \cF_{\inter}^-} \prod_{i=1}^{i_f} M^{1} \Big]\,,\crcr
 &\leq& K'^{L} K''^{F_{\inter}}\Big[\prod_{\ell \in \cL}\prod_{(i,k)\in \N^2/ \ell \in G^k_i} M^{-1}\Big]
 \Big[\prod_{f \in \cF_{\inter}^+} \prod_{(i,k)\in \N^2/ l_f \in G^k_i} M^{2} \Big]\Big[
\prod_{f \in \cF_{\inter}^-} \prod_{(i,k)\in \N^2/ l_f \in G^k_i} M^{1} \Big]\,,
\eea
where $l_f$ is the strand in $f$ such that $i_{l_f}=i_f=\min_{\ell \in f}i_\ell$. 
Then, we have
\bea
|A_\cG| &\leq&  K'^{L} K''^{F_{\inter}}\Big[\prod_{(i,k)} \;\prod_{\ell \in \cL(G^k_i)}M^{-1}\Big]
 \Big[\prod_{(i,k)} \prod_{f \in \cF_{\inter}^+\cap G^k_i;\, l_f \in f \cap G^k_i} M^{2} \Big]\Big[
\prod_{(i,k)} \prod_{f \in \cF_{\inter}^-\cap G^k_i;\, l_f \in f \cap G^k_i} M^{1} \Big]\,.
\eea
However, the set $\{l_f \in f \cap G^k_i\}$ is empty whenever 
$ f \cap G^k_i$ is an open face of $G^k_i$. Recall that $G^k_i$
contains only line with index $i_\ell \geq i$ and that $i_{\ell_f}$
is the smallest index among the strand indices of $f$. 
Thus, if $ f \cap G^k_i$ happens to be open then $i_{l_f}<i$
which cannot occur in $G^k_i$. It is straightforward to 
obtain
\bea
|A_\cG| &\leq&  K'^{L} K''^{F_{\inter}}\Big[\prod_{(i,k)} M^{-L(G^k_i)}\Big]
 \Big[\prod_{(i,k)} \prod_{f \in \cF_{\inter}^+(G^k_i)} M^{2} \Big]\Big[
\prod_{(i,k)} \prod_{f \in \cF_{\inter}^-(G^k_i)} M^{1} \Big]\crcr
&\leq & K''' \prod_{(i,k)} M^{-L(G^k_i)+2F^+_{\inter}(G^k_i)+F^-_{\inter}(G^k_i)} \,,
\label{powe}
\eea
where $L(G^k_i)= |\cL(G^k_i)|$ and $F^\pm_{\inter}(G^k_i)= |\cF_{\inter}^\pm(G^k_i)|$, and $K'''$ a constant including all
other constants. 

We now understand the main feature introduced
the field redefinition. 
Given graph $\cG$, the $T3$ model generates a superficial power-counting yielding a divergence degree, forgetting a moment mass and wave function
counter-terms, $\omega_d(\cG) = -L(\cG) + F_\inter(\cG)$, where
$L(\cG)$ is the number of lines and $F_\inter(\cG)$ the number
of closed strands or internal faces of the graph $\cG$ \cite{BenGeloun:2012pu}. 
After mapping $T3 \to \sigma T3$,
one realizes that, from \eqref{ci},
 each line $l$ at scale $i_l$ provides a convergent
factor of $M^{-i_l}$ yielding, roughly,  a convergent factor
of $M^{-i L(\cG)}$. This is exactly similar to the situation of the $T3$ model.
In contrast, the number of internal faces
has been drastically reduced by merging the strands after the rank reduction. There
are two types of internal faces: those indexed by $p_f$ in $\cF_\inter^-(\cG)$  
which yield a divergent factor $M^{i}$ \eqref{sum2};
and internal faces labeled by $\sqrt{n_f}$ 
in $\cF_\inter^+(\cG)$ 
yielding a greater divergent factor of $M^{2i}$ from \eqref{sum1}. 
The faces in $\cF_\inter^+(\cG)$ bring twice the contribution of their
analogues in $\cF_\inter(\cG)$ in $T3$. This compensate the lost contribution of the missing faces after the tensor to matrix reduction and, from that, one recovers the balanced power-counting of the initial model.

One notices that the analysis of the amplitude has been performed
without taking into account the wave function counter-terms. 
These are two-point graphs formed with two-leg vertices $V_{2;s}$, $s=1,2,$  related to \eqref{wfct}. 
Including these in the above analysis is simple and from  
equation \eqref{powe}, we are able to identify an initial power counting theorem 

\begin{theorem}[Power counting] For a connected graph $\cG$
(with external arguments integrated on test functions)
the amplitude is bounded by 
\beq
|A_\cG| \leq  K^n \prod_{(i,k)\in \N^2} M^{\omega_d(G^k_i)} \,,
\eeq
where $K$ and $n$ are large constants and the divergence
degree of any graph $G$ is given by 
\beq
\omega_d(G) = -L(G)+ V_{2,1}(G) + V_{2,2}(G) +2F_{\inter}(G)\,.
\label{degdiv}
\eeq
\end{theorem}
\proof From \eqref{powe}, the sole remaining point is simply solved by 
\beq
2F^+_{\inter}(G)+F^-_{\inter}(G) \leq 2 F_{\inter}(G)\,.
\eeq

\qed

The bound by $\omega_d(G) \leq -L(\cG) + 2 F_\inter(\cG)$
is, in fact, optimal since there are some configurations where $F_\inter(\cG) = F_\inter^+(\cG)$. 

\subsection{Divergence degree analysis}
\label{subsect:div}

\noindent{\bf Divergence degree in topological terms.}
In order to re-express the divergence degree of a graph 
in term of its topological components, we first need 
to introduce the  notion of ``pinching'' of an open graph or a graph with external legs. 
This notion has been initially defined for an arbitrary rank $D$ colored tensor graph in the reference  \cite{Gurau:2009tz}. Here, we  apply the pinching procedure for a ribbon graph with external legs. 

\begin{figure}[h]
\begin{minipage}[t]{.8\textwidth}
\begin{center}
\includegraphics[height=2cm]{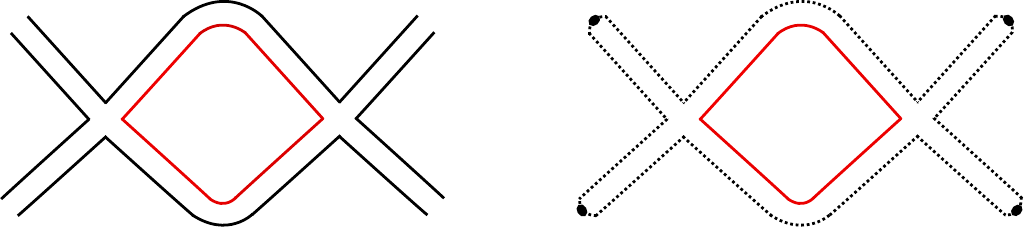}
\hspace{1cm}
\includegraphics[height=1.2cm]{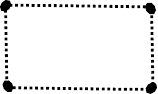}
\vspace{5mm}
\caption{ {\small An open ribbon graph $\cG$, its pinched version
$\widetilde\cG$ and boundary $\bG$. An closed face $f_1$ (in red) present both in  $\cG$ and $\widetilde\cG$
and an additional face $f_2$ (in dot) only present in $\widetilde\cG$.}
\label{fig:pinch}}
\end{center}
\end{minipage}
\put(-326,-11){$\cG$}
\put(-180,-11){$\widetilde\cG$}
\put(-65,-11){$\bG$}
\put(-338,25){$f_1$}
\put(-228,25){$f_2$}
\put(-193,25){$f_1$}
\end{figure}

The pinching procedure of a ribbon graph $\cG$ with external legs consists in the construction of another graph $\widetilde\cG$ by the gluing of 2-leg vertices at each external leg of $\cG$. See Figure \ref{fig:pinch} for an illustration. After pinching a open ribbon graph $\cG$, one obtains a closed
ribbon graph $\widetilde\cG$, with the same set of vertices and set of edges but a different set of faces. There are two types of faces in $\widetilde\cG$, those coming
from $\cG$ (see $f_1$ in Figure \ref{fig:pinch}) and others coming from the external faces of $\cG$ (see $f_2$ in Figure \ref{fig:pinch}). Hence
\beq
\cF_{\inter}(\widetilde\cG) = \cF_{\inter}(\cG) \cup 
\cF'_{\ext} \,,\qquad 
|\cF_{\inter}(\widetilde\cG)| = F_{\inter}(\cG) + 
|\cF'_{\ext}|\,. 
\label{factg}
\eeq

Associated with the pinching, there exists another underlying graph. 
We define the boundary graph $\bG$ of $\cG$, the closed graph 
with set of vertices given by the 2-leg vertices introduced 
during the pinching and set of lines given by the set $\cF_{\ext}(\cG)$
of external faces  of $\cG$ (see an illustration in Figure \ref{fig:pinch}). Therefore, 
\beq
V(\bG) = N_{\ext}(\cG)\,, \quad
L(\bG) = F_{\ext}(\cG). 
\eeq
There is no difficulty to infer that 
\beq
\cF'_{\ext}  = C_{\bG} \,. 
\label{fexc}
\eeq
where $C_{\bG}$ is the number of connected components
of $\bG$. 

\begin{proposition}[Divergence degree] Let  $\cG$ be  a connected
graph. The divergence degree of $\cG$ is 
\beq
\omega_d(\cG) =-V_2 -\frac12\big[ N_{\ext} - 4\big]
- 4 g_{\widetilde\cG} - 2(C_{\partial\cG} -1)\,,
\label{degdivtopo}
\eeq
where $V_2$ is the number of mass  renormalization vertices,
$g_{\widetilde\cG}$ is the genus of the closed pinched graph
$\widetilde\cG$ extending the initial graph $\cG$, $C_{\partial\cG}$ is the number of connected components
of $\partial\cG$ the boundary graph associated with $\cG$ 
and $N_{\ext}$ is the number of external legs of $\cG$.
\end{proposition}
\proof Let $\cG$ be a ribbon graph of the theory. 
It has $V_4(\cG)$ number of vertices of the $\varphi^4$ type,  
$V_{2,0}(\cG)$, number of vertices of mass renormalization, 
and $V_{2,s=1,2}(\cG)$, 
number of vertices of wave function renormalization,  
$L(\cG)$ number of internal 
lines, $F_{\inter}(\cG)$ number closed faces, $F_{\ext}(\cG)$
number of open faces. The following expression relates as, in any $\varphi^4$ theory, the number of vertices and lines:
\bea
&&
V(\cG) = V_4(\cG) + \sum_{s=0,1,2} V_{2,s}(\cG)\,, \crcr
&&
4V_4(\cG) + 2 \sum_{s=0,1,2} V_{2,s}(\cG) = 2L(\cG) + N_{\ext}(\cG)\,, \qquad  L(\cG) - 2V_4(\cG) =\sum_{s=0,1,2} V_{2,s}(\cG)  - \frac12 N_{\ext}(\cG)\,.
\label{verline}
\eea
Consider now the pinched graph $\widetilde \cG$ and its
Euler characteristics: 
\beq
2 - 2g_{\widetilde \cG} = V(\cG) - L(\cG) + F_{\inter}(\widetilde \cG)\,,
\label{euler}
\eeq
where $g_{\widetilde \cG}$ denotes the genus of $\widetilde \cG$. 
Combining both \eqref{factg} and \eqref{fexc}, we infer from \eqref{euler}
that
\bea
F_{\inter}(\cG) = 2 - 2g_{\widetilde \cG} - [V(\cG) - L(\cG) + C_{\bG}]\,.
\eea
The degree of divergence of $\cG$ \eqref{degdiv} therefore can be rewritten as
\bea
\omega_d(\cG)  &=& -L(\cG) + 4- 4g_{\widetilde \cG} - 2[V(\cG) - L(\cG) + C_{\bG}] + \sum_{s=1,2} V_{2,s}(\cG) \crcr
&=& 4- 4g_{\widetilde \cG}  -V_{2,0}(\cG)- \frac12 N_{\ext}(\cG) -2 C_{\bG}\,,
\eea
where, in the last step, we used \eqref{verline}. Hence \eqref{degdivtopo}
is immediate. 
\qed

\medskip 

\noindent{\bf Primitively divergent graphs.}
From \eqref{degdivtopo}, one realizes that the marginal 
 or log-divergent graphs with $N_{\ext}=4$ are determined by 
\beq
\omega_d(\cG) =0: \qquad \quad 
g_{\widetilde \cG} =0, \quad  C_{\bG} =1,
\quad V_2=0\,.
\label{4pt}
\eeq
They should correspond to a vertex renormalization. 
For $N_{\ext}=2$, there are two categories of divergent graphs: 
\bea
&&
\omega_d(\cG)=1: \qquad \quad 
g_{\widetilde \cG} =0, \quad C_{\bG} =1,\quad 
V_2=0\,, \crcr
&&\omega_d(\cG) =0: \qquad \quad 
g_{\widetilde \cG} =0,\quad C_{\bG} =1,\quad V_2=1\,.
\label{2pt}
\eea
 These should contribute to a mass 
and wave function renormalizations.

A remarkable fact is that all these divergent contributions
are planar graphs. We remark also that this power-counting is in agreement with the one of the GW model in 4D for which the renormalization procedure
identifies as relevant graphs only planar graphs with one broken external face with at most four external legs \cite{Grosse:2004yu}.

There is  now a clear effect induced by the projection map.
In the renormalization procedure of $T3$ \cite{BenGeloun:2012pu}, the dominant
contributions, at  given $N_\ext=4,2$, were specific graphs called
``melonic'' \cite{Bonzom:2011zz} with melonic
boundary graph $\bG$ having a unique connected component. 
In the present context, one realizes the fact that dominant contributions should be melonic has been washed
away and gets replaced just by a planarity condition.
This can be indeed expected since the new model is a
matrix theory which can only be sensitive to an ordinary planarity condition.
Nevertheless, for melonic graphs $F^+_{\inter}$ should
be the one which contributes so that one
would rather have a divergence degree much more
restrictive and of the form $\widetilde\omega_d(\cG)  = -L(\cG) + 2 F^+_\inter(\cG)$ so that many planar graphs which would be
not melonic would be simply cast away for being convergent
if one use such a power counting theorem.

\subsection{Renormalization}
\label{subsect:renorm}

We can now study the subtraction terms and provide the proof that
$N_\ext$-point functions expand in  divergent parts of the form of the initial Lagrangian terms plus a convergent remainder. 
It is at this stage that one finds important that 
the kinetic term is of form $ap^\alpha+bn^\beta$
in order to obtain the correct wave function counter-terms.

\medskip 

\noindent{\bf Renormalization of the four-point functions.}
Consider a 4-point quasi local subgraph $G^k_i$ as determined by the conditions \eqref{4pt}  with four external propagators. The external momenta of this graph should follow the pattern of the vertex as given by 
Figure \ref{fig:vertex2d}. We shall denote the external 
momenta associated to each of the four external faces $f^-_x$ and $f^+_x$, $x=1,2$, by  $p_{f^-_x}$, $n_{f^+_x}$.  
We write $q^-_{f,x}= p_{f^-_x}$ and 
$q^+_{f,x}= \sqrt{n_{f^+_x}}$. 
These external momenta are at scale $j\ll i$. 
To internal faces $f^\pm$, correspond internal 
momenta $p_{f^-}$ or $n_{f^+}$ such that, using a single momentum variable, we write $q^-_{f}=p_{f^-}$ and $q^+_{f}=\sqrt{n_{f^+}}$. 
Internal momenta are of scale $i_f\geq i \gg j$.
We also introduce the notation $a_{+}= b$ and $a_-=a$. 

The propagator lines (both internal and external) of the graph are indexed by $\ell$. An external line will be
particularly indexed by $l$.  The amplitude of $G^k_i$ writes (in loose notations, we drop the dependence in $G^k_i$, from now)
\bea
A_4[\{q^\pm_{f,x}\}] =  \sum_{q^\pm_{f,i}} 
\int [\prod_{\ell }d\alpha_\ell e^{-\alpha_\ell\mu}] \;\prod_{\epsilon=\pm}\Big[\prod_{f\in F^\epsilon_{\ext}} e^{-[\sum_{\ell \in f} \alpha_\ell] a_\epsilon q^\epsilon_{f,x}}\Big]
\Big[\prod_{f\in F^\epsilon_{\inter}} e^{-[\sum_{\ell \in f} \alpha_\ell]  a_\epsilon q^\epsilon_{f} }\Big] ,
\eea
where $\alpha_\ell \in [M^{-i_\ell+1}, M^{i_\ell}]$. For an external propagator then $ \alpha_l \in [M^{-j_l+1}, M^{j_l}]$ with $j_l \ll i$.  In the product involving $f\in F^\epsilon_{\inter}$,
all $\alpha_\ell$ are in the slice $i_\ell \geq i \gg j_l$.

Expanding the amplitude of one external face as a part 
involving external $\alpha_l$ and internal $\alpha_{\ell \neq l}$,
for any $t\in [0,1]$, we define a parameterized amplitude as
\beq
A_4[\{q^\pm_{f,x}\};t] =  \sum_{q^\pm_{f,i}} 
\int [\prod_{\ell }d\alpha_\ell e^{-\alpha_\ell\mu}] \;\prod_{\epsilon=\pm}\Big[\prod_{f\in F^\epsilon_{\ext}}e^{-(\alpha_l + \alpha_{l'}) a_\epsilon q^\epsilon_{f,x}}
e^{-t(\sum_{\ell \in f/\ell \neq l} \alpha_\ell)a_\epsilon  q^\epsilon_{f,x}}\Big]
\Big[\prod_{f\in F^\epsilon_{\inter}} e^{-[\sum_{\ell \in f} \alpha_\ell]  a_\epsilon q^\epsilon_{f} }\Big] ,
\eeq
which interpolates between $A_4[\{q^\pm_{f,x}\};t=1] = A_4[\{q^\pm_{f,x}\}]$ and $A_4[\{q^\pm_{f,x}\},t=0]=A_4[\{0\}]$. 
The next stage is to perform a Taylor expansion around $t=0$ 
of $A_4[\{q^\pm_{f,x}\};t=1]$  which correspond to an expansion of the amplitude around the local part of the graph $G^i_k$. 
We have 
\beq
A_4[\{q^\pm_{f,x}\};t]\Big|_{t=1} = A_4[\{q^\pm_{f,x}\};t]\Big|_{t=0}
+ \int_0^1 \frac{d}{dt} A_4[\{q^\pm_{f,x}\};t] \,dt\,.
\eeq 
Then, at zeroth order, we get the amplitude
\beq
A_4[\{q^\pm_{f,x}\};0] =\sum_{q^\pm_{f,i}} 
\int [\prod_{\ell }d\alpha_\ell e^{-\alpha_\ell\mu}] \;\prod_{\epsilon=\pm}
\Big(\Big[\prod_{f\in F^\epsilon_{\ext}}e^{-(\alpha_l + \alpha_{l'}) a_\epsilon q^\epsilon_{f,x}}\Big] \Big[ \prod_{f\in F^\epsilon_{\inter}} e^{-[\sum_{\ell \in f} \alpha_\ell]  a_\epsilon q^\epsilon_{f} }\Big] \Big)\,.
\eeq
Scrutinizing the term depending on the external momenta $q^\pm_{f,x}$,
we find that this factorizes and yields
\bea
A_4[\{q^\pm_{f,x}\};0] &=&
C_{j_1}([q^-_{f,1},q^+_{f,1}]) C_{j_2}([q^-_{f,2},q^+_{f,1}])
C_{j_3}([q^-_{f,2},q^+_{f,2}]) C_{j_4}([q^-_{f,1},q^+_{f,2}])\times
\crcr
&&
\sum_{q^\pm_{f,i}} 
\int [\prod_{\ell \neq l }d\alpha_\ell e^{-\alpha_\ell\mu}] \;\prod_{\epsilon=\pm}
\Big(  \prod_{f\in F^\epsilon_{\inter}} e^{-[\sum_{\ell \in f} \alpha_\ell]  a_\epsilon q^\epsilon_{f} } \Big)
\eea
This amplitude is log-divergent and corresponds to 
a graph with 4 propagator lines the external 
data of which coincide with a vertex of the model. 
In other word, $A_4[\{q^\pm_{f,x}\};0]$ contributes
to the vertex renormalization. 

Focusing on the remainder, we have the following bound
\bea
|R_4|
& = & \Big| \int_0^1 \Big\{\sum_{q^\pm_{f}} 
\int [\prod_{\ell }d\alpha_\ell e^{-\alpha_\ell\mu}] \;
\sum_{\epsilon=\pm;\, f\in F^\epsilon_{\ext}}
\Big[-[\sum_{\ell \in f/\ell \neq l} \alpha_\ell] a_\epsilon q^\epsilon_{f,x}
\Big] 
\prod_{\epsilon=\pm}\Big(\Big[\prod_{f\in F^\epsilon_{\ext}}
e^{-[\sum_{\ell \in f/\ell \neq l} \alpha_\ell] a_\epsilon t q^\epsilon_{f,x}}e^{-(\alpha_l+ \alpha_{l'})  a_\epsilon q^\epsilon_{f,x} }\Big]\crcr
&& \times\qquad 
\Big[\prod_{f\in F^\epsilon_{\inter}}
 e^{-[\sum_{\ell \in f} \alpha_\ell]  a_\epsilon q^\epsilon_{f} }\Big] \Big)\Big\}dt\Big| \crcr
&\leq&\Big|\Big\{\sum_{q^\pm_{f}} 
\int [\prod_{\ell }d\alpha_\ell e^{-\alpha_\ell\mu}] \;
\sum_{\epsilon=\pm;\, f\in F^\epsilon_{\ext}}
\Big[-[\sum_{\ell \in f/\ell \neq l} \alpha_\ell]  (\alpha_l+ \alpha_{l'})^{-1}
\Big]
\prod_{\epsilon=\pm}\Big(\Big[\prod_{f\in F^\epsilon_{\ext}}
 e^{-\frac12(\alpha_l+ \alpha_{l'})  a_\epsilon q^\epsilon_{f,x} }\Big]\crcr
&& \times 
 \qquad 
\Big[\prod_{f\in F^\epsilon_{\inter}}
 e^{-[\sum_{\ell \in f} \alpha_\ell]  a_\epsilon q^\epsilon_{f} }\Big] \Big) \int_0^1 e^{-\sum_{\epsilon=\pm;f\in F^\epsilon_{\ext}}[\sum_{\ell \in f/\ell \neq l} \alpha_\ell] a_\epsilon t q^\epsilon_{f,x}}dt\Big\}\Big| \crcr
&&
\eea
where, passing from the equality to the inequality, we use 
$
A |X| e^{-A|X|} \leq e^{-A |X|/2} 
$
with $|X|=q^\epsilon_{f,x}$ and  $A =a_\epsilon(\alpha_l+\alpha_{l'})$. 
We can optimize the bound on $|R_4|$ by choosing 
\beq
 e(G^k_i) = \sup_{l\in G^k_i} j_l \,,\quad \alpha_l \geq M^{- e(G^k_i)}\,;
\qquad
i(G^k_i) = \inf_{\ell \in G^k_i} i_\ell\,,\quad 
\alpha_{\ell} \leq M^{- i(G^k_i)}\,,\;\; \ell \neq l\,,
\label{estim}
\eeq
such that there exists some constant $c$ such that
\bea
|R_4|&\leq&c M^{-(i(G^k_i)-e(G^k_i))}\Big|\Big\{\sum_{q^\pm_{f}} 
\int [\prod_{\ell }d\alpha_\ell e^{-\alpha_\ell\mu}] \;\prod_{\epsilon=\pm}\Big( \Big[\prod_{f\in F^\epsilon_{\ext}}e^{-\frac12(\alpha_l+ \alpha_{l'})  a_\epsilon q^\epsilon_{f,x} }\Big]
\Big[\prod_{f\in F^\epsilon_{\inter}}
 e^{-[\sum_{\ell \in f} \alpha_\ell]  a_\epsilon q^\epsilon_{f} }\Big] \Big)\crcr
&& \times 
 \qquad  \int_0^1 e^{-\sum_{\epsilon=\pm;f\in F^\epsilon_{\ext}}[\sum_{\ell \in f/\ell \neq l} \alpha_\ell] a_\epsilon t q^\epsilon_{f,x}}dt\Big\}\Big| .
\eea
The integral in $t$ yields $O(1)$ as well as integrals over external $\alpha_l$. 
At the end, up to some constant $K$, one gets 
\bea
|R_4| \leq K M^{-(i(G^k_i)-e(G^k_i))} M^{-\omega_d(G^k_i)=0}\,.
\eea
The factor $M^{-(i(G^k_i)-e(G^k_i))}$ brings an additional decay ensuring 
both an improved power counting 
and the final summability over the scale attribution $\mu$ \cite{Rivasseau:1991ub}.  

\medskip 

\noindent{\bf Renormalization of the two-point functions.}
We now focus on 2-point quasi local subgraphs $G^k_i$ as given by  \eqref{2pt} with two external propagators. We will however not treat
both cases but only the linearly divergent graph. The logarithmically
divergent contribution can be recovered by simple inference. 

The external momenta of the graph should follow the pattern of 
mass vertex with two external 
momenta associated to each of the two external faces $f^-_0$ and $f^+_0$,  by  $p_{f^-_0}$, $n_{f^+_0}$.  As above, we use a compact script $q^-_{0}= p_{f^-_0}$ and 
$q^+_{0}= \sqrt{n_{f^+_0}}$. 
We use the same conventions and notations for internal and external scales
as done before. 

The amplitude of $G^k_i$ writes 
\beq
A_2[\{q^+_0,q^-_{0}\}] =  \sum_{q^\pm_{f,i}} 
\int [\prod_{\ell }d\alpha_\ell e^{-\alpha_\ell\mu}] \;\Big[\prod_{\epsilon=\pm} e^{-[\sum_{\ell \in f^\epsilon} \alpha_\ell] a_\epsilon q^\epsilon_{0}}\Big]
\prod_{\epsilon=\pm}\Big[\prod_{f\in F^\epsilon_{\inter}} e^{-[\sum_{\ell \in f} \alpha_\ell]  a_\epsilon q^\epsilon_{f} }\Big] ,
\eeq
where $\alpha_\ell \in [M^{-i_\ell+1}, M^{i_\ell}]$ and, for external propagators, $ \alpha_l \in [M^{-j_l+1}, M^{j_l}]$ with $j_l \ll i$.

Using again a parameterized amplitude 
\bea
A_2[\{q^+_0,q^-_{0}\};t] =  \sum_{q^\pm_{f,i}} 
\int [\prod_{\ell }d\alpha_\ell e^{-\alpha_\ell\mu}] \;\Big[\prod_{\epsilon=\pm} e^{-(\alpha_l + \alpha_{l'}) a_\epsilon q^\epsilon_{0}}
e^{-t[\sum_{\ell \in f^\epsilon/\ell \neq l} \alpha_\ell] a_\epsilon q^\epsilon_{0}} \Big]
\prod_{\epsilon=\pm}\Big[\prod_{f\in F^\epsilon_{\inter}} e^{-[\sum_{\ell \in f} \alpha_\ell]  a_\epsilon q^\epsilon_{f} }\Big] ,
\eea
providing an interpolation from $A_2[\{q^+_0,q^-_{0}\};t=1]$
to $A_2[\{q^+_0,q^-_{0}\};t=0]$, we perform a Taylor expansion 
such that
\bea
A_2[\{q^+_0,q^-_{0}\};t]\Big|_{t=1} = A_2[\{q^+_0,q^-_{0}\};t]\Big|_{t=0}
+ \frac{d}{dt} A_2[\{q^+_0,q^-_{0}\};t]\Big|_{t=0} +
\int_0^1 (1-t)\frac{d^2}{dt^2} A_2[\{q^+_0,q^-_{0}\};t] \,dt\,.
\label{intay}
\eea 
We focus on the first term:
\bea
 A_2[\{q^+_0,q^-_{0}\};0]&=&\int [\prod_{l}d\alpha_l e^{-\alpha_l\mu}] \Big[\prod_{\epsilon=\pm} e^{-(\alpha_l + \alpha_{l'}) a_\epsilon q^\epsilon_{0}} \Big]
\sum_{q^\pm_{f,i}} 
\int [\prod_{\ell \neq l}d\alpha_\ell e^{-\alpha_\ell\mu}] \;
\prod_{\epsilon=\pm}\Big[\prod_{f\in F^\epsilon_{\inter}} e^{-[\sum_{\ell \in f} \alpha_\ell]  a_\epsilon q^\epsilon_{f} }\Big] \crcr
&=&
C[q^-_0, q^+_0]\, C[q^-_0, q^+_0] \,
\sum_{q^\pm_{f,i}} 
\int [\prod_{\ell \neq l}d\alpha_\ell e^{-\alpha_\ell\mu}] \;
\prod_{\epsilon=\pm}\Big[\prod_{f\in F^\epsilon_{\inter}} e^{-[\sum_{\ell \in f} \alpha_\ell]  a_\epsilon q^\epsilon_{f} }\Big]
\eea
which is linearly divergent and, clearly, renormalizes  the mass 
in the model. 

The second term can be written in the  form
\bea
 A_2'[\{q^+_0,q^-_{0}\};0] &=&  \sum_{q^\pm_{f,i}} 
\int [\prod_{\ell }d\alpha_\ell e^{-\alpha_\ell\mu}] \;
\sum_{\epsilon=\pm}\Big[-[\sum_{\ell \in f^\epsilon/\ell \neq l} \alpha_\ell] a_\epsilon q^\epsilon_{0}\Big]
\Big[
\prod_{\epsilon=\pm}e^{-(\alpha_l + \alpha_{l'}) a_\epsilon q^\epsilon_{0}} \Big]
\prod_{\epsilon=\pm}\Big[\prod_{f\in F^\epsilon_{\inter}} e^{-[\sum_{\ell \in f} \alpha_\ell]  a_\epsilon q^\epsilon_{f} }\Big] \crcr
&=& \int [\prod_{l }d\alpha_l e^{-\alpha_l\mu}] \Big[
\prod_{\widetilde\epsilon=\pm}e^{-(\alpha_l + \alpha_{l'}) a_{\widetilde\epsilon} q^{\widetilde \epsilon}_{0}} \Big] \times
 \crcr
&&\sum_{\epsilon=\pm}\Bigg\{a_\epsilon q^\epsilon_{0}
\sum_{q^\pm_{f,i}} 
\int [\prod_{\ell\neq l }d\alpha_\ell e^{-\alpha_\ell\mu}] \;
\Big[-\sum_{\ell \in f^\epsilon/\ell \neq l} \alpha_\ell\Big]
\prod_{\epsilon'=\pm}\Big[\prod_{f\in F^{\epsilon'}_{\inter}} e^{-[\sum_{\ell \in f} \alpha_\ell]  a_{\epsilon'} q^{\epsilon'}_{f} }\Big] \Bigg\}
\eea
This term contribute to the wave function renormalization and should  separately provide two log-divergent contributions to $a_+\sqrt{n}$ and
to $a_-p$ of the initial kinetic term. Indeed, 
the first integral in $d\alpha_l $ yields again two factorized external propagators
and the integral in $d\alpha_{\ell \neq l}$ yields a logarithmic divergence
by the power counting. This is clearly the case since the sum 
$\sum_{\ell \in f^\epsilon/\ell \neq l} \alpha_{\ell} \leq  k M^{-i_{\ell_0}}
= k\prod_{(i,k)/ \ell_0 \in G^k_i}M^{-1}$, lowers the linear divergence of one unit. 

It is definitely at this stage that the introduction of two 
wave function couplings, namely $a_\pm$, plays a drastic role
by providing the necessary freedom to renormalize wave functions
with different coefficients. Indeed, due to the fact that 
the model is no longer symmetric in strands, it is not
true that the contributions to the wave function renormalization of $a_+$
might find corresponding contributions in the wave function
renormalization for $a_-$. As a simple illustration, at one-loop,
consider the unique tadpole graphs $T^+$ and $T^-$  given in Figure
\ref{fig:tad} in the appendix which are involved in the self-energy. At this order of perturbation,  $T^+$ contribute to the wave function renormalization 
associated with $a p$. Meanwhile the contribution of $T^-$
becomes finite. There is therefore, at this order and at any higher order, 
no term restoring the symmetry between 
for the wave function renormalizations  $ap\delta_{Z_1}$ and $b \sqrt{n}\delta_{Z_2}$.

The last term needed to be analyzed in \eqref{intay} is the remainder
\bea
|R_2| &=&   \Big| \int_0^1 (1-t) dt \sum_{q^\pm_{f,i}} 
\int [\prod_{\ell }d\alpha_\ell e^{-\alpha_\ell\mu}] \; 
\Big(
\sum_{\epsilon=\pm}\Big[-[\sum_{\ell \in f^\epsilon/\ell \neq l} \alpha_\ell] a_\epsilon q^\epsilon_{0}\Big]\Big)^2
\crcr
&& 
\Big[\prod_{\epsilon=\pm} e^{-(\alpha_l + \alpha_{l'}) a_\epsilon q^\epsilon_{0}}
e^{-t[\sum_{\ell \in f^\epsilon/\ell \neq l} \alpha_\ell] a_\epsilon q^\epsilon_{0}} \Big]
\prod_{\epsilon=\pm}\Big[\prod_{f\in F^\epsilon_{\inter}} e^{-[\sum_{\ell \in f} \alpha_\ell]  a_\epsilon q^\epsilon_{f} }\Big]
\Big| \crcr
&\leq&
   \Big| \int_0^1 (1-t) dt \sum_{q^\pm_{f,i}} 
\int [\prod_{\ell }d\alpha_\ell e^{-\alpha_\ell\mu}] \; 
\Big(
\sum_{\epsilon=\pm}\Big[-[\sum_{\ell \in f^\epsilon/\ell \neq l} \alpha_\ell]\Big](\alpha_l+\alpha_{l'})^{-1}\Big)^2
\crcr
&& 
\Big[\prod_{\epsilon=\pm} e^{-(\alpha_l + \alpha_{l'}) a_\epsilon q^\epsilon_{0}/4}
e^{-t[\sum_{\ell \in f^\epsilon/\ell \neq l} \alpha_\ell] a_\epsilon q^\epsilon_{0}} \Big]
\prod_{\epsilon=\pm}\Big[\prod_{f\in F^\epsilon_{\inter}} e^{-[\sum_{\ell \in f} \alpha_\ell]  a_\epsilon q^\epsilon_{f} }\Big]
\Big| 
\eea
where,  passing to the inequality, we used twice the 
bound $A q^\epsilon e^{-A q^\epsilon } \leq e^{-A q^\epsilon /2}$. 
Then using again the best estimates \eqref{estim}, 
we can find an optimal bound on $|R_2|$ as, up to a constant $c$, 
\bea
|R_2| &\leq&
 c M^{-2(i(G^k_i)-e(G^k_i))}   \Big|  \sum_{q^\pm_{f,i}} 
\int [\prod_{\ell }d\alpha_\ell e^{-\alpha_\ell\mu}] \; 
\prod_{\epsilon=\pm}\Big( \Big[e^{-(\alpha_l + \alpha_{l'}) a_\epsilon q^\epsilon_{0}/4} \Big]\Big[\prod_{f\in F^\epsilon_{\inter}} e^{-[\sum_{\ell \in f} \alpha_\ell]  a_\epsilon q^\epsilon_{f} }\Big]\Big)\crcr
&&\times
\int_0^1 (1-t) dt \Big[\prod_{\epsilon=\pm} e^{-t[\sum_{\ell \in f^\epsilon/\ell \neq l} \alpha_\ell] a_\epsilon q^\epsilon_{0}} \Big]
\Big| .
\eea
The integrals in $t$ and in $d\alpha_l$ yield a mere $O(1)$ factor.
Thus, one proves that this remainder $|R_2|$ is bounded and provides
an additional decay of $M^{-2(i(G^k_i)-e(G^k_i))}$ to $M^{\omega_d(G^k_i)=1}$ which will ensure the final
 summability on momentum attribution $\mu$ in the standard way of
\cite{Rivasseau:1991ub}.

This achieves the proof of the renormalizability of the model
and hence Theorem \ref{theo1} holds.

\subsection{RG flow of wave function renormalizations and a super-renormalizable vector model}
\label{subsect:vect}

The $T3$ model with symmetric interactions has been proved asymptotically
free \cite{BenGeloun:2012pu}. The projection on the matrix model 
proves to preserve that behavior and the model $\sigma T3$
is also asymptotically free. The proof of this statement has been provided in the
appendix. There is however more that we can say about the latter
situation. 

One notices that two wave function renormalizations should be introduced  at this level 
\beq
Z_a = 1- \frac1a \partial_{p}\Sigma(p,\sqrt{n})|_{p=0=n}\,,
\qquad 
Z_b = 1- \frac1b \partial_{\sqrt{n}}\Sigma(p,\sqrt{n})|_{p=0=n}\,,
\eeq 
where $\Sigma(p,\sqrt{n})$ is the so-called self-energy 
or sum of all two-point one-particle irreducible (1PI) contributions. 
In the appendix, we discuss at first order how behave
these terms.  At one-loop computations, one finds that the contribution to $\frac1b \partial_{\sqrt{n}}\Sigma(p,\sqrt{n})|_{p=0=n}$ is finite (see the appendix). 
In contrast, 
$Z_a$ is a log-divergent quantity and should be used 
to renormalize $a$. It becomes immediate that, the renormalized
quantity associated with $b$, namely $b^\ren$, is such that    
$b^\ren =b(Z_b/Z_a)^{1/2}\to 0$. 
Thus, the model flows towards a free model with vanishing $b$ in its kinetic term.

\begin{figure}[h]
\begin{minipage}[t]{.8\textwidth}
\begin{center}
\includegraphics[height=1.3cm]{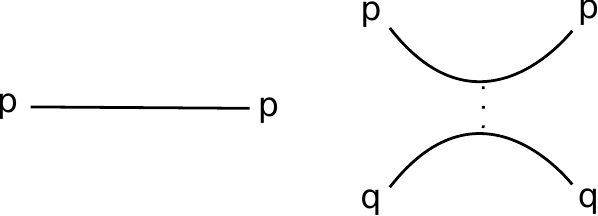}
\hspace{5mm}
\caption{ {\small Propagator and vertex $v_4$ of the vector model. }
\label{fig:vect}}
\end{center}
\end{minipage}
\end{figure}

Having this remark in mind, one could ask if the above model 
does not imply the existence of another renormalizable model flowing towards the same Gaussian fixed point with vanishing $b$. This question is certainly non trivial. 
The simplest way to think about this is to start building a model with the same kinetic term as for the $\sigma T3$ but with vanishing $b$. We furthermore need to gauge fix the sector $n$ in all field $\phi_{p,n}$, otherwise, there will be a free mode sum in this sector. Using, for instance,  the $n=0$ mode sector, 
one builds a the kinetic term and interaction of the form 
\bea
S^{\kin, 0}= \sum_{p}\bar\phi_{p,0}(a p + \mu) \phi_{p,0}\,,
\qquad 
S^{\inter, 0} = \left(\sum_{p} \bar\phi_{p,0}\phi_{p,0}\right)^2
\label{newkin}
\eea
which describes, surprisingly, a vector model. Propagator and vertices 
can be  represented by Figure \ref{fig:vect}.

We can perform a multi-scale analysis in a similar way that was introduced in Subsection \ref{subsect:multi}. It is simple to obtain this limit case since
it correspond to set $b=0$ in the analysis therein. One gets propagator 
bound given by \eqref{ci} at $b=0$ and, 
for any connected graph $\cG$ and for a momentum scale attribution $\mu$,
an optimal bound on amplitude as
\bea
|A_{\mu}| \leq K^n \, \prod_{(i,k)} M^{\omega_d(G^i_k)}\, ,
\qquad
\omega_d(G) = - L(G^i_k) + F_{\inter}(G^i_k)\,,
\eea
where, once again, closed strands are the sources of divergences. 

\begin{figure}[h]
\begin{minipage}[t]{.8\textwidth}
\begin{center}
\includegraphics[height=1cm]{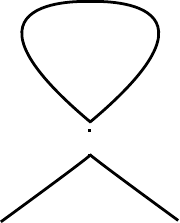}
\hspace{5mm}
\caption{ {\small Tadpole graph. }
\label{fig:tadvect}}
\end{center}
\end{minipage}
\end{figure}

Since the vertex \eqref{newkin} is apparently disconnected from 
the point of view of its external legs (it is clearly a factor of two  
pieces), we introduce $v'_2$ made with two 
external legs and being half of this vertex $v_4$. Consider now
a graph $\cG$ connected with respect to only vertices of the form $v'_2$. 
Call $V'$ the number of vertices of this graph, $L$ its number of lines
and $F_{\inter}$ is number of closed loops, $C$ its number of 
connected components, $N_{\ext}$ its number of external legs.
We have the relations
\bea
2V' = 2L + N_{\ext}\,, \qquad 
C= F_{\inter} + C'\,,
\eea
where $C'$ counts the number of external strands. 
For $\cG$, it is immediate to translate the above divergence degree 
 as
\beq
\omega_d(\cG) =  -L + F_{\inter} = -V' +\frac12 N_{\ext}+ C-C' \,.
\eeq
Having assumed that $\cG$ is connected,
then $C=1$. A connected component in the theory is either
an open strand or a closed one, then either $N_{\ext}=2$, $C'=1$, $F_{\inter}=0$
or $N_{\ext}=0$, $C'=0$ and $F_{\inter}=1$. In all cases, the divergence
degree recasts as 
\bea
\omega_d(\cG) = -V' + C\,.
\eea
From this, we can give the list of all primitively divergent graphs
which reduces to a unique type of graph:
\bea
C=1\,, \quad \omega_d(\cG) =  -V' +1 = 0 \; \Leftrightarrow \; 
(V'=1,\; L=1,\; N_{\ext} = 0) \,.
\eea
This corresponds to a unique 1-loop graph which is logarithmically divergent.
Coming back to the situation with graphs with vertices $v_4$, the same type of graphs is nothing but  a tadpole graph (see Figure \ref{fig:tadvect}) which should only involve a mass renormalization. Thus this model is super-renormalizable. Clearly, this feature is similar to a ordinary scalar $\varphi^2_2$.

\section{New classes of renormalizable models
from previous models}
\label{sect:nclass}

In this section, we use the above mechanism to reveal
the existence of new renormalizable models issued from
well-known  renormalizable tensor and matrix models.

\subsection{Classes of GW models}
\label{subsect:gw}

The GW model is the first discovered renormalizable model pertaining to both matrix models and noncommutative geometry \cite{Grosse:2004yu}\cite{Grosse:2003nw}.
This model proves to be renormalizable at all orders by
curing a previous undesirable effect called UV/IR
mixing affecting renormalization procedure on noncommutative
spaces \cite{Rivasseau:2007ab}. The UV/IR mixing is simply removed by adding an harmonic term of the
form $\Omega^2 \tilde x^2$, where $\Omega$ is an harmonic
frequency, $\tilde x = 2(\Theta^{-1})_{\mu\nu} x^\mu$,
with the noncommutative structure given by $[x^\mu,x^\nu]= \Theta^{\mu\nu}$, $x^\mu \in \R^D$, $D=2,4$.

We will restrict the study to complex fields and will
place ourselves at the self-dual point
$\Omega =1$ for which, in the continuum, the kinetic term
of the GW model is $(-\Delta + \tilde x^2 + \mu)$, $\mu$
being some IR mass regulator, so that
the model becomes dual in momenta and positions.
In 4D, the complex GW model is given by the  action
(Euclidean signature) \cite{Grosse:2004yu}
\bea
S_{GW;4D} = \int d^4x\;
\Big\{ \frac12 \bar\phi(x)(-(\partial_\nu)^2 + \tilde x^2 + \mu)\phi(x)
 + \frac{\lambda}{4} \bar\phi(x)\star \phi(x) \star\bar\phi(x) \star \phi(x) \Big\}\,,
\label{actionGW4d}
\eea
where $\star$ denotes the Moyal star product.
There exists a basis for which the above model
finds another clear  translation. This is the so-called  matrix basis \cite{Grosse:2004yu} where 
each field can be viewed as a rank 4 complex tensor $\phi_{\vec p,\vec q}$ where $\vec p = (p_1,p_2)$ and  $\vec q =(q_1,q_2) \in \N^2$.  The same GW action reads as, at the self-dual point $\Omega=1$, 
\bea
S_{GW;4D} = \frac12\sum_{\vec p, \vec q \in \N^2}
\bar\phi_{\vec p, \vec q} \Big[ |p| +|q|  + \mu \Big] \phi_{\vec q, \vec p } +
\frac{\lambda}{4}
\sum_{\vec m, \vec n, \vec p, \vec q \in \N^2}
\bar\phi_{\vec m, \vec n}\,\phi_{\vec n, \vec p}\,
\bar\phi_{\vec p, \vec q}\,\phi_{\vec q, \vec m}\,.
\label{gw4d}
\eea
where we introduce the notation, for any $\vec n \in \N^2$, $|n| =n_1 + n_2$. 
Call this the $GW_{4D}$ model.
The propagator and vertex of this model can be represented
as in Figure \ref{fig:vertex2d} but one should regard each strand
as doubled. The interaction is clearly of the cyclic form so that 
the procedure introduced above applies naturally here. 

Let us first recall few facts about the renormalizability of the $GW_{4D}$ \cite{Grosse:2004yu}.
The propagator in the slice $i$ admits the bound, for some 
constant $K$, 
\bea
C_i(\vec p ;\vec q) \leq K M^{-i} e^{-M^{-i}( |p| + |q|+ \mu)}\,.
\eea
After a multi-scale analysis, one is led to a power counting 
theorem giving the divergence degree of any connected
graph $\cG$ as (up to unessential two-point vertices which
can be contracted)
\beq
\omega_d(\cG) = -L(\cG) + 2 F_{\inter}(\cG) = -4g(\widetilde\cG) - 2(C_\partial(\cG) - 1) 
- \frac12(N_{\ext}(\cG) - 4)\,,
\label{grwpow}
\eeq
using similar notations as in previous section. 
Thus only open ribbon graphs $\cG$ characterized with 4 and 2 external legs, a vanishing genus of their pinching $g(\widetilde\cG)=0$ and a unique connected component
of the boundary should be the ones inducing divergent 4- and 2-point functions. 4-pt graphs should participate to the coupling constant $\lambda$ renormalization and 2-pt graphs should involve mass
and wave function renormalizations.  

We now start the program dealing with the rank reduction of the model. 
The $\sigma$ projection leads us to a new
rank 3 model, called $\sigma GW_{4D}$, described by the following action:
\beq
S_{\sigma GW;4D}= \frac12\sum_{p\in N,  \vec q \in \N^2}
\bar\phi_{p, \vec q} \Big[a\sqrt{p}+b|q|  +\mu\Big] \phi_{\vec q,p} +
\frac{\lambda}{4} \sum_{m, p \in \N;\; \vec n,\vec q \in N^2}
\bar\phi_{m, \vec n}\,\phi_{\vec n, p}\,
\bar\phi_{p, \vec q}\,\phi_{\vec q, m}\,,
\label{s1gw4d}
\eeq
where we introduce wave coupling parameters, $a$
and $b$, in order to have {\it a priori} a proper notion of wave function renormalizations. Note that, in the new action, the mapping
$\sigma$ can be applied as well on the second couple
of integers $\vec q$.  Hence, combinatorially,
we have two such $\sigma GW_{4D}$ models.

The propagator of the model \eqref{s1gw4d} is given by 
\beq
C(\{p,\vec q\,\};\{\tilde p, \vec{\tilde{q}}\,\}) =\delta_{p,\tilde p}\delta_{\vec q,\vec{\tilde q}} /(a\sqrt{p}+b|q|  +\mu)\,.
\eeq
Introducing a UV cut-off on momenta, 
we claim that this model is just renormalizable at all orders
of perturbations. We shall sketch the main phases of the proof
of this statement since the details of all arguments can be fully recovered from Section \ref{sect:rank3}. 

First, we bound the propagator kernel in the slice $i$ as
\bea
C_i(p,\vec q)\leq K M^{-i} e^{-M^{-i}(a\sqrt{p}+ b|q|  +\mu)}\,.
\label{propbound}
\eea
Performing the multi-scale analysis using this sliced propagator bound, it is immediate to realize that, open faces do not
participate to the power-counting and that there is two types 
of closed faces: faces parameterized by $\sqrt{p}$ 
which have a double weight and faces with momentum 
$q_1$ or $q_2$. These latter faces go always by pairs
$(q_1,q_2)$. The closed face amplitudes evaluation yields:   
\beq
\sum_{p} e^{-a M^{-i} \sqrt{p}} = \frac{2}{a^2}  M^{2i} (1+ O(M^{-i})) \,,
\qquad 
\sum_{q_1,q_2} e^{-b M^{-i} (q_1+q_2)} = \frac{1}{b^2}  M^{2i} (1+ O(M^{-i})) \,. 
\label{faceamp}
\eeq
Thus \eqref{faceamp} shows that, even though
the number of faces of the reduced theory certainly decreases in $p$-sector, the amplitude of each face becomes twice greater in that
sector. This will be the key feature ensuring again the renormalizability of $\sigma GW_{4D}$.

The degree of divergence of any connected graph $\cG$ becomes, in
the same anterior notations (and conventions forgetting two-point vertices): 
\beq
\omega_d(\cG) = - L(\cG)  + 2F^+_{\inter}(\cG)  +  2F^-_{\inter}(\cG) 
= -L(\cG) + 2F_{\inter}\,.
\eeq
Since the arguments yielding \eqref{grwpow} depend uniquely on the topology of graphs, and provided the topology of graphs of $GW_{4D}$ and of graphs of $\sigma GW_{4D}$ is the same, one ends up with the same degree of divergence. As a consequence, the list of divergent
graphs are identical for both models. The renormalization of 
4-pt and 2-pt functions can be checked in the way 
of Subsection \ref{subsect:renorm}. However, 
another interesting point which has to be fully inspected is the possibility of having a unique wave function renormalization for
$\sigma GW_{4D}$.

Interestingly, we can prove that  putting $a=b\sqrt{2}$, the new model is still renormalizable\footnote{Note that this simply means that we could have introduced a different choice of action $S_{\sigma GW;4D}$ \eqref{s1gw4d} with a kinetic term such that 
\beq
 \frac12\sum_{p\in N,  \vec q \in \N^2}
\bar\phi_{p, \vec q} \Big[a\sqrt{2p}+ b(q_1 +q_2)  +\mu\Big] \phi_{\vec q,p}\;,
\eeq
well motivated by the $\sigma$ map. In this case, 
one  sets $a=b$ and still get the renormalizability
for the subsequent model. 
}.
This can be viewed as follows. Using a Taylor expansion of a general two-point function around its ``local'' divergence (see corresponding paragraph in Subsection \ref{subsect:renorm}), setting $a=b\sqrt{2}$, we must prove that the divergent contribution associated with the kinetic terms $\sqrt{2p}$ and
 $(q_1+ q_2)$ are the same. This point will ensure that a unique wave function renormalization $\delta_Z \sim \log \Lambda$ can be defined as $b\delta_Z(\sqrt{2p} + q_1+ q_2)$. This statement can be verified explicitly from \eqref{faceamp} with now $a=b\sqrt{2}$ and using the fact that the symmetry of the strands present in the model \eqref{gw4d} is in fact preserved in the model \eqref{s1gw4d}. Indeed,
as previously claimed, the pair of faces parameterized by $(q_1,q_2)$
can be labeled by a unique $\vec q$ and can be merged. 
Whenever one has a contribution to the wave function renormalization
$b\sqrt{2p}\,\delta_Z$, we can find a symmetric partner with a similar
divergence renormalizing $b|q|\,\delta_Z$. Hence, 
both actions with $a\neq\sqrt{2}b$ or $a=\sqrt{2}b$ define rank 3 renormalizable rank 3 tensor models. 

As a consequence, in all families of GW models discussed below, 
there exist still another reduction obtained by fixing $a\propto b$  
(depending on the sector where the $\sigma$ mapping is
applied) leading again to a renormalizable reduced model.

Applying another reduction in the remaining sector $\vec q$,  we get
finally a matrix model that we call $\sigma^2 GW_{4D}$
given by
\beq
S_{\sigma^2 GW;4D}= \frac12\sum_{p,  q \in \N}
\bar\phi_{p, q} \Big[a\sqrt{p}+ b\sqrt{q}  +\mu\Big] \phi_{p, q} +
\frac{\lambda}{4} \sum_{m, n, p, q \in \N}
\bar\phi_{m, n}\,\phi_{n, p}\,
\bar\phi_{p, q}\,\phi_{q, m}\,.
\label{s2gw4d}
\eeq
The model $\sigma^2 GW_{4D}$ is strand symmetric and can be studied along the lines of the anterior analysis. From the first equation of \eqref{faceamp}, there are two types of faces coined by $\sqrt{p}$ or by $\sqrt{q}$ with equal weight  $M^{2i}$. Following step by step, the above procedure, we are led to the same power counting theorem
and the same type of graphs which ought to be renormalized. 
The renormalization of these graphs can be performed
as earlier done. It is immediate that the model $\sigma^2 GW_{4D}$  is renormalizable at all orders. Note that we can put now $a=b$ involving
the possibility of having a unique wave function renormalization. This is
 without consequence on the renormalizability property of the model due to the restored symmetry of all strands in the propagator.

It is noteworthy that $S_{\sigma^2 GW;4D}$ which is a rank 2 model 
does not correspond to the action of the GW model in 2D denoted
$GW_{2D}$. The $GW_{2D}$ model is super-renormalizable and 
is given by the action \cite{Grosse:2003nw}:
\bea
S_{GW;2D} = \frac12\sum_{p,  q \in \N}
\bar\phi_{p, q} \Big[p+q + \mu \Big] \phi_{p, q} +
\frac{\lambda}{4}\sum_{m, n, p, q \in \N}
\bar\phi_{m, n}\,\phi_{n, p}\,
\bar\phi_{p, q}\,\phi_{q, m}\,.
\label{actGW2D}
\eea
The propagator and vertex of this model match with Figure \ref{fig:vertex2d}. 
The proposition of the super-renormalizability of this model 
can be quickly reviewed. The propagator in a slice meets 
a bound
\beq
C_i(\{p,q\};\{\tilde p, \tilde q\}) \leq K M^{-i} e^{-M^{-i}(p+q + \mu)}\,.
\eeq
From this bound and a multiscale analysis of the
amplitude of a connected graph $\cG$, we can write an optimal bound amplitude associate with $\cG$ with degree of divergence
(with similar anterior conventions)
\beq
\omega_d(\cG)= - L(\cG)  + F_{\inter}(\cG)
= -2g_{\widetilde\cG} - (V(\cG)-1) - (C_\partial(\cG)-1)\,,
\label{ome2D}
\eeq
where one should use the Euler characteristics to map 
the different numbers. In this form, one realizes that 
the more the graph contains vertices the more it is 
convergent, a specific feature of super-renormalizability. 
The list of primitively divergent graphs summarizes as
follows: since $V\geq 1$ and $C_\partial(\cG)\geq 1$
(we do not discuss vacuum graphs $C_\partial(\cG)=0$), 
the only possibility for $\omega_d(\cG)\geq 0$ reads
\beq
g_{\widetilde\cG}=0\,,\quad V(\cG)=1 \,, \quad C_\partial(\cG)=1 
\qquad \omega_d(\cG)=0\,.
\eeq
This is nothing but a tadpole graph yielding a log-divergent
contribution which re-absorbed by a mass renormalization.

The action \eqref{actGW2D} is already in the matrix form. 
We can consider now the reversed of the $\sigma$ process. 
Reshuffling now \eqref{actGW2D} by using the inverse $\sigma^{-1}$
in one sector, say $q$ without loss of generality,
$\sigma^{-1}(q)= \vec q$, we get a rank 3 GW model:
\bea
S_{GW;2D} =S'_{GW;2D}&=&
\frac12\sum_{p\in \N, \vec q \in \N^2}
\bar\phi_{ p, \vec q} \Big[ ap + b\sigma(\vec q) + \mu \Big] \phi_{ \vec q,p} +
\frac{\lambda}{4}
\sum_{m, p \in \N;\, \vec n, \vec q \in \N^2}
\bar\phi_{m, \vec n}\,\phi_{\vec n, p}\,
\bar\phi_{ p, \vec q}\,\phi_{\vec q, m}
\label{cs1gw}
\eea
which can be related to the new $\sigma^{-1} GW_{2D}$ model
defined as
\bea
S_{\sigma^{-1}GW;2D}=
\frac12\sum_{p\in \N, \vec q \in \N^2}
\bar\phi_{p, \vec q} \Big[ ap +b(q^2_1 + q^2_2) + \mu \Big] \phi_{ p, \vec q} +
\frac{\lambda}{4}
\sum_{m, p \in \N;\;\vec n, \vec q \in \N^2}
\bar\phi_{m, \vec n}\,\phi_{\vec n,p}\,
\bar\phi_{p, \vec q}\,\phi_{\vec q, m}\,.
\label{s1gw2d}
\eea
One can check that, using \eqref{boundN}, that the propagators
 \eqref{cs1gw} and \eqref{s1gw2d} have same behavior
in a slice. We have the propagator for $\sigma^{-1}GW_{2D}$
given by 
\beq
C(\{p,\vec q\,\};\{\tilde p,\vec{\tilde q}\,\})
= \delta_{p,\tilde p}\delta_{\vec q,\vec{\tilde q}}/ (ap +b(q^2_1 + q^2_2) + \mu)\,.
\eeq
We must show that this model is again super-renormalizable
using a momentum cut-off and the recipe by now used. 

Bounding the propagator in a slice gives
\beq
C_i(\{p,\vec q\,\};\{\tilde p,\vec{\tilde q}\,\}) \leq 
K M^{-i}e^{-M^{-i}(ap +b(q^2_1 + q^2_2) + \mu)}\,. 
\eeq
The multi-scale analysis leads us to consider
two types of faces. One labeled by $ap$ and a
pair of faces always labeled by $\vec q$. The closed face amplitude evaluation associated with these new faces yields:   
\beq
\sum_{q_1,q_2} e^{- \delta M^{-i} b(q^2_1+q^2_2)} = \frac{\pi}{4\delta^2 b^2}  M^{i} (1+ O(M^{-i/2}))\,. 
\label{faceamp2}
\eeq
In this specific instance, the fact that the number of faces
in $q$-sector is increasing is merely compensated by the fact that each face will be associated  with a less divergent factor of $M^{-i/2}$. The ensuing power counting is the same as the one 
determined by \eqref{ome2D}, yielding the same type of divergent
graphs involved only in the mass renormalization. Thus, the $\sigma^{-1}GW_{2D}$ is super-renormalizable. 

Discussing the possibility of merging wave function couplings, since the
$\sigma^{-1}GW_{2D}$ model  is super-renormalizable with only mass renormalization, there is actually no point to merge or not 
couplings $a$ and $b$. A symmetric model will be however 
the one which has two tadpoles with exactly the same amplitude.
This can be provided by the identification $a=4b^2/\pi$ in \eqref{s1gw2d}.

Applying now $\sigma^{-1}$ in the $p$-sector,
we can infer that the following rank 4 GW model  which
will be referred to as $\sigma^{-2}GW_{2D}$:
\bea
S_{\sigma^{-2}GW;2D}&=&
\frac12\sum_{\vec p, \vec q \in \N^2}
\bar\phi_{\vec p, \vec q} \Big[ a(p^2_1 + p^2_2) +b(q^2_1 + q^2_2) + \mu \Big] \phi_{\vec p, \vec q} +
\frac{\lambda}{4}
\sum_{\vec m, \vec n, \vec p, \vec q \in \N^2}
\bar\phi_{\vec m, \vec n}\,\phi_{\vec n, \vec p}\,
\bar\phi_{\vec p, \vec q}\,\phi_{\vec q, \vec m}\,.
\label{s2gw2d}
\eea
The proof that this model is super-renormalizable
can be easily performed according to the previous
case. The interesting point is again that doubling the faces
in each sector $\vec p$ or $\vec q$, is still controlled by the fact 
each of the new face amplitude is less divergent and behaves like
$M^{i/2}$. This maintains the balance and allows us to recover the same super-renormalizable power counting theorem. 

It can be asked the continuum
models underlying  \eqref{s1gw4d}, \eqref{s2gw4d}, \eqref{s1gw2d} and
\eqref{s2gw2d} and their relation to noncommutative geometry. At this point, an answer to that question is not clear. One can investigate the particular forms of the propagators which might lead to other
interesting kinetic terms extending the ordinary $(p,x)$-duality
which has led to the control of UV/IR mixing. However, as
explained at the beginning, these actions might be useful 
in another  context of nonlocal field theories called 
Tensorial Group Field Theory (TGFT) \cite{Rivasseau:2011hm} 
different from noncommutative field theory on Moyal spaces.
Indeed, in \cite{BenGeloun:2011rc}, a rank 4 tensor model extending
the above $T3$ tensor model has been proved to
be renormalizable at all order of perturbation theory.
Fields $\varphi: U(1)^4 \to \C$ can be viewed as rank four tensors $\varphi_{p_1,p_2,p_3,p_4}$, $p_i \in \Z$. The kinetic part
of this model is given by  closely related to the kinetic
part of \eqref{s2gw2d}. However, the interactions of these two models are different. In \cite{BenGeloun:2011rc}, one type of interaction is cyclic
(hence can be recast in a matrix form) and  another cannot be recast in terms of matrix trace. This makes this particular higher rank TGFTs non trivial with this respect but definitely susceptible to simplified using the above analysis.

\subsection{Families of renormalizable tensor models}

By iterating the procedure, we can generate  three
different families of models related
either to the GW models or to the $T3$ model. We establish that, for the three models,
\bea
{\textrm{$T3$ Class}}: &&\qquad  \dots \to \sigma^{-n} T3 \to \dots  \to \sigma^{-1} T3 \to T3 \to \sigma T3 \,;
\cr\cr
{\textrm{$GW_{4D}$ Class}}: &&\dots \to \sigma^{-n} GW_{4D} \to \dots  \to \sigma^{-2} GW_{4D}\to \sigma^{-1} GW_{4D} \to  GW_{4D} \to \sigma GW_{4D} \to \sigma^2 GW_{4D} \,;
\cr\cr
{\textrm{$GW_{2D}$  Class}}: &&\qquad  \dots \to \sigma^{-n} GW_{2D} \to \dots  \to \sigma^{-2} GW_{2D}\to \sigma^{-1} GW_{2D} \to GW_{2D} \,.
\eea
Note that each arrow might lead to different theories according
to the choice of indices on which the reduction
or extension are performed. For instance,  $T3 \to \sigma T3 $
leads to a unique model whereas $GW_{4D} \to \sigma GW_{4D} \to \sigma^2 GW_{4D}$ leads to two models and
$\sigma^{-2} GW_{2D} \leftarrow \sigma^{-1} GW_{2D} \leftarrow GW_{2D}$
leads as well to two models.
A way to classify all these models might be to consider
as belonging to the same family or class those
having a common and initial matrix model.

We claim that all models issued from $GW_{4D}$
and $\sigma T3$ are just renormalizable and all models from 
$GW_{2D}$ are super-renormalizable. The justification of this 
has been in fact already established. Let us formalize that proof
in full generality for the sake of clarity. 

Let $\cG$ be a connected graph in any of the above 
model $\sigma^{n}(\cdot)$. Call $\sigma\cG$ and $\sigma^{-1}\cG$  the graphs corresponding to $\cG$ in $\sigma^{n+1}(\cdot)$ and 
$\sigma^{n-1}(\cdot)$, respectively. Both are uniquely 
defined by $\cG$. $\sigma\cG$ is obtained from $\cG$ after merging two faces corresponding to the collapse of two collated (or cyclically disposed) strand momenta in one, i.e.
$a(p_1^\alpha + p_2^\alpha) \to a p^{\alpha/2}$. Meanwhile, $\sigma^{-1}\cG$ is obtained after splitting of one strand momentum into two cyclic ones in $\cG$, say $a p^\alpha \to a(p_1^{2\alpha} + p_2^{2\alpha})$. 
The proof or our claim rests on the  following statement:

\begin{lemma}[Stability of degree of divergence]
Let $\cG$ be a connected graph in any of the above 
model $\sigma^{n}(\bullet)$, $\bullet= GW_{4D}, T3,GW_{2D}$. 
Let $\omega_{d,n,\bullet}(\cG)$ the degree of divergence of $\cG$
in $\sigma^{n}(\bullet)$. Then 
\bea
\omega_{d,n,\bullet}(\cG) =  \omega_{d,n\pm 1,\bullet}(\sigma^{\pm 1}\cG)\,.
\eea 
\end{lemma}\proof  Given a graph $\cG$, the multi-scale analysis yields, uniquely from propagator bounds, the quantity $\prod_{(i,k)} M^{-L(G^i_k)}$. This contribution is identical for all models. 
For $\sigma^n(\bullet)$, let us call  $F_{\inter,n}(\cG)$ the number of faces of $\cG$. This number divides as 
\beq
F_{\inter,n}(\cG) = F'_{\inter, n}(\cG) + F_{\inter,n}^{\alpha}(\cG)\,,
\eeq
where $F_{\inter,n}^{\alpha}(\cG)$ counts uniquely the number
of faces associated with a particular strand momentum $p^\alpha$, or two cyclic momenta such that $(p^\alpha_1, p^\alpha_2)$. 
Once the model is fixed, for a given graph, $F_{\inter,n}^{\alpha}$ is known. Applying $\sigma^{\pm 1}$ on $\cG$, we have
\beq
 F_{\inter, n}' (\cG)=  F_{\inter, n \pm 1}' (\sigma^{\pm 1}\cG)\,, 
\qquad 
F_{\inter,n}'(\cG) = 2^{\pm 1} F_{\inter,n\pm 1}'(\sigma^{\pm 1}\cG)\,.
\label{fprim}
\eeq
The stability of the power counting theorem results from the following facts: For the process $\sigma^{-1}$ (resp. $\sigma$): the splitting  of one closed face in two produces at the same time a reduction of  the divergence of the face amplitudes by half (resp. the merging of two faces in one increases by a factor of two the divergence of that latter face). In the initial $\sigma^{n}(\bullet)$, consider a face with a momentum $p^\alpha$. Then the amplitude of this closed face is, for large $i$, 
\beq
\sum_{p} e^{-\delta M^{-i} a p^{\alpha} } =c M^{\frac i\alpha}
(1+ O(M^{-\frac i\alpha}))\,, 
\eeq
up to some constant $c$. After applying $\sigma^{-1}$
(resp. $\sigma$), we obtain from one face two collated faces (resp.
from two collated faces one face) with  amplitude
\beq
\sum_{p_1} e^{-\delta M^{-i} b p_1^{2\alpha}}
= c' M^{\frac{i}{2\alpha}}
(1+ O(M^{-\frac{i}{2\alpha}}))\,
\qquad 
\text{(resp. }
\sum_{p_1} e^{-\delta M^{-i} b p_1^{\frac{\alpha}{2}}}
= c' M^{\frac{2i}{\alpha}}
(1+ O(M^{-\frac{2i}{\alpha}}))\,\text{ )}\,,
\eeq
 for some constant $c'$. Thus, for any cases, bearing in mind \eqref{fprim}, the optimal bound amplitude gives a degree of divergence,
for some constant $\beta$, 
\bea
\omega_{d,n, \bullet}(\cG)&=& 
-L(G^i_k) +  \frac{1}{\alpha} \frac{2}{2} F^{\alpha}_{\inter; n}(G^k_i)  
+ \beta F'_{\inter; n}(G^k_i)  \cr\cr
&=& -L(G^i_k)
+ \frac{1}{\alpha}2^{\mp} F^{\alpha}_{\inter; n\mp 1}(G^k_i) +\beta F'_{\inter; n\mp 1}(G^k_i)  = \omega_{d,n\mp 1, \bullet}(\sigma^{\mp 1}\cG)\,.
\eea

\qed

The above proposition shows that the degree of divergence does not depend on $n$. The proof of renormalizability in a particular class  holds because, all the models in that class have an identical power counting theorem leading to an identical list of primitively divergent
graphs with the same degree of divergence. The subtraction scheme
and renormalization of each diverging $N$-point function can be undertaken in 
a standard way. Note that for the $\sigma^n GW_{4D}$ and $\sigma^n T3$ models, one may require to adjust properly the $a$'s and $b$'s
in order to have a well defined wave function renormalization
if the initial model is defined with a single wave function coupling.  
Otherwise, one may always insert new wave function couplings after each
momentum splitting in order to define as much as wanted wave function renormalizations. This leads us to the following proposition.

\begin{theorem}[Classes of renormalizable models]\label{theo3}
The models $\sigma^{n}GW_{4D}$, $n\in (-\infty, 2]$, 
$\sigma^{n}T3$, $n\in (-\infty, 1]$,  are just renormalizable
at all orders. 
The models $\sigma^{n}GW_{4D}$, $n\in (-\infty, 0]$
are super-renormalizable at all orders. 
\end{theorem}

Being interested in the change of the propagator for these
different theories, we have the following table:

\vspace{0.5cm}
\begin{center}
\begin{tabular}{  llllllllll }
Matrix &&&&&Rank 3&&&&Rank 4\\
 \hline  \hline  \\
$\sigma^2 GW_{4D}:$  $\sqrt{p} +\sqrt{q}$&
&&&& $\sigma GW_{4D}:$  $a\sqrt{p} +b(q_1 +q_2)$
&&&& $GW_{4D}:$  $p_1+p_2+q_1 +q_2$
\\
&\\
$\sigma T3:$  $ap +b\sqrt{q}$
&&&&& $T3:$  $ap +b(q_1+q_2)$
 &&&& $\sigma^{-1} T3:$  $a(p^2_1 +p_2^2) +b(q_1+q_2)$ \\
&\\
$GW_{2D}:$  $p +q$&&&&&
$\sigma^{-1}GW_{2D}:$  $ap +b(q_1^2+q_2^2)$
&&&&
$\sigma^{-2}GW_{2D}:$  $p_1^2 + p_2^2 +q_1^2+q_2^2$\\
&\\
 \hline\hline
\end{tabular}
\end{center}

\section{Outlook}

The present work affords a link between different
nonlocal renormalizable theories using tensor fields of rank $\geq 2$
appearing in different field theory contexts (noncommutative field theory and  tensor models).
It provides also a machinery in order to generate
classes of models sharing the significant
property to be renormalizable. 

Remarkably,  the renormalizability of the $T3$ model can
be understood from the fact that,
its multi-scale analysis can be performed in the reduced
matrix model $\sigma T3$. For the $\sigma T3$ model, the propagator
in a given slice $i$ contains $\sqrt{q}$ which allows to
have a similar power counting of $\sigma^2 GW_{4D}$
(the renormalizability of which holds from the 
renormalizability of $GW_{4D}$). One also should emphasize that studying the $T3$ model equipped with a single interaction naturally forced us to break the symmetry between strand indices in that interaction term. It results from this consideration that the propagator should possess different wave function  couplings $a$ and $b$ in order to have well-defined wave function renormalizations. In the UV regime, this explicit symmetry breaking  therefore leads to a peculiar free model with
kinetic term having vanishing in part. By simple inference, within the set  of our basic axioms, we build a toy model corresponding to this kinetic term. The said model turns out to be a super-renormalizable vector model.

As additional insights of the above study, let us comment that, since $GW_{2D}$ is super-renormalizable \cite{Grosse:2003nw}, by introducing a set of
propagators  
\beq
C^{\epsilon}[\{p,q\};\{\tilde p,\tilde q\}]
= \delta_{p,\tilde p} \delta_{q,\tilde q}/(ap+bq^{\frac12+\epsilon} +\mu )\,,
\qquad \epsilon \in [0,1/2]\,,
\eeq
we find a continuum of theories interpolating between $GW_{2D}$ for $\epsilon=1/2$ and $\sigma T3$ for $\epsilon=0$, all with the property
from being super-renormalizable to being just-renormalizable. Indeed, 
following step by step our analysis, one finds a divergence degree
for a connected graph $\cG$ as
\bea
\omega_{d, GW_{2D}}(\cG)\leq \omega_d^{\epsilon}(\cG) = - L(\cG) + F^-_{\inter;\epsilon}(\cG) + \left(\frac{2}{1+2\epsilon}\right)F^+_{\inter;\epsilon}(\cG) \leq \omega_{d, T3}(\cG)\,,
\eea
leading us to the fact that only tadpoles with internal face
momenta $p$ or $q^{\frac12+\epsilon}$ diverge. Pursuing the propagator interpolation and using
the propagator 
\beq
C^{\epsilon'}[\{p,q\};\{\tilde p,\tilde q\}]
= \delta_{p,\tilde p} \delta_{q,\tilde q}/(ap^{\frac12+\epsilon'}+bq^{\frac12}+\mu)\,,
\qquad \epsilon' \in [0,1/2]\,,
\eeq
we find another continuum of theories being all just-renormalizable
 leaving from $\sigma  T3$ at $\epsilon'=1/2$
to $\sigma^2 GW_{4D}$ at $\epsilon'=0$. Indeed, in this situation, we have another divergence degree given by 
\bea
\omega_{d,  T3}(\cG)\leq \omega_d^{\epsilon'}(\cG) = - L(\cG) +\left(\frac{2}{1+2\epsilon'}\right) F^-_{\inter;\epsilon'}(\cG) + 2F^+_{\inter;\epsilon'}(\cG)\leq \omega_{d,  GW_{4D}}(\cG)\,.
\eea
At any $\epsilon'\in [0,1/2]$, $\omega_{d,  GW_{4D}}(\cG)$ is a upper bound which can be saturated by $\omega_d^{\epsilon'}(\cG)$. The reasoning becomes the same as for the proof of the renormalizability of $\sigma T3$. In any cases, the divergence subtraction scheme will remain the same. The $T3$ model is, with this respect, ``critical''.

One may also ask about the UV behavior of these classes
of models. It can be shown that, all the models such that
$ap^{\frac12 + \epsilon} +bq^{\frac12}$, for $\epsilon\in ]0,1/2]$,
are asymptotically free in the UV (see the appendix for a proof
of this claim) and, at the end-point $\epsilon=0$, the model
becomes safe which corresponds, of course, of the
well-known asymptotic safeness of the GW model in 4D \cite{Grosse:2004yu}\cite{Rivasseau:2007ab}.

Finally, it would be interesting to provide a space-time representation to all these matrix/tensor theories, the same way that the GW model is defined by the action \eqref{actionGW4d}
as a field theory living on non-commutative $\R^4$. Since terms in $\sqrt{p}$ in the propagator seem slightly awkward to translate as differential operators (introducing another type of nonlocality), it seems more natural to work with the theories with linear propagator, for instance $p_1+p_2+q_1+q_2$ for  $GW_{4D}$ and $ap+b(q_1+q_2)$ for $T3$. Considering $T3$, one can write it naturally as a group field theory \cite{oriti} on $U(1)^3$ with the kinetic term given by the sum of the derivative with respect to each coordinate. Written as such, we lose a priori the relation with noncommutative field theory. However, one could similarly write the  $GW_{4D}$ model as a group field theory on $U(1)^4$. From this perspective, it seems interesting to investigate  in the future the relationship between non-commutative field theories of the Moyal-type and group field theories. A possible approach could be to push further the relation between the Moyal star-product and the non-commutativity based on group manifolds as investigated in \cite{NCSU2}.

\section*{Acknowledgements}
Discussions with Razvan Gurau are gratefully acknowledged.
Research at Perimeter Institute is supported by the Government of Canada through Industry
Canada and by the Province of Ontario through the Ministry of Research and Innovation.

\section*{Appendix: One-loop $\beta$-function of the $( ap^{\frac12+\epsilon}+bq^{\frac12})$--model}
\appendix

\renewcommand{\theequation}{A.\arabic{equation}}
\setcounter{equation}{0}

We prove in this appendix that all the models with
propagators of the form $ap^{\frac12 + \epsilon} +bq^{\frac12}$, $\epsilon \in ]0,1/2]$ are asymptotically free in the UV.
The calculation of the $\beta$-function is made for $\epsilon=1/2$ corresponding to the $\sigma T3$ model.
For the remaining models, the proof is totally similar.

The $\beta$-function of a $\phi^4$ theory is generally encoded in the ratio
\bea
\lambda^{\text{ren}} = - \frac{\Gamma_4(0,0,0,0)}{Z^2} \,,
\label{lamren}
\eea
where $\lambda^{\text{ren}}$ is the renormalized
coupling (and so $\lambda$ stands for the bare coupling).  $\Gamma_4(m,n,p,q)$ is  the sum of amputated 1PI 
four-point functions truncated at one-loop and which should 
be computed at zero external momenta in \eqref{lamren}. $Z$ is the wave function renormalization
which should involve the subleading log-divergent term
obtained after the Taylor expansion of the self-energy
$\Sigma$ which is the sum of the 1PI amputated two-point
functions truncated at one-loop.

\begin{figure}[h]
\begin{minipage}[t]{.8\textwidth}
\begin{center}
\includegraphics[height=2cm]{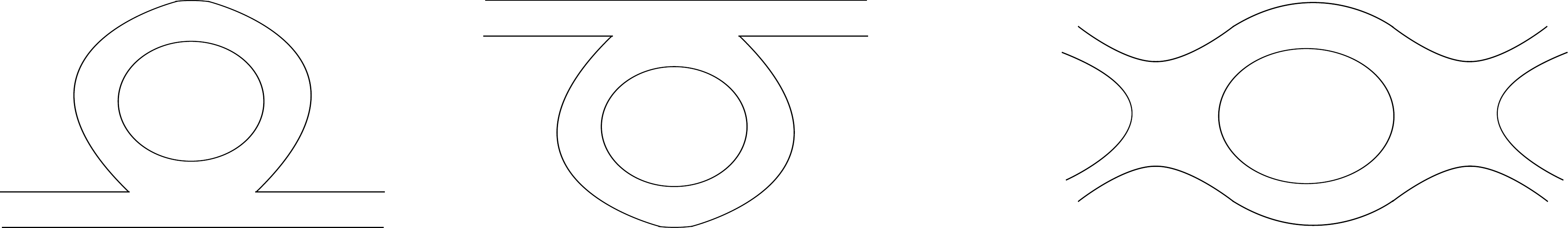}
\hspace{5mm}
\caption{ {\small Tadpoles up $T^+$ and down $T^-$ and 
four-point graph $F$. }
\label{fig:tad}}
\end{center}
\end{minipage}
\put(-360,-20){$T^+$}
\put(-400,10){$m$}
\put(-360,30){$\sqrt{k}$}
\put(-400,-8){$\sqrt{n}$}
\put(-280,58){$m$}
\put(-235,25){$k$}
\put(-280,40){$\sqrt{n}$}
\put(-235,-20){$T^-$}
\put(-130,55){$m$}
\put(-80,25){$\sqrt{k}$}
\put(-140,30){$\sqrt{n}$}
\put(-130,0){$p$}
\put(-75,-20){$F$}
\end{figure}

Coming back to our present model, 
$\Sigma$  involves two types of contributions
called tadpoles ``up'' $T^+$ and ``down'' $T^-$(see Fig.\ref{fig:tad}).

We have at one-loop 
\beq
\Sigma =A_{T^+} + A_{T^-}=
\frac{(-\lambda)}{2} \Big[ 2 \sum_{k} \frac{1}{am + b\sqrt{k} + \mu}
+ 2\sum_{k} \frac{1}{ak + b\sqrt{n} +  \mu}\Big]
+ O(\lambda^2)\,.
\eeq
As noticed, $\Sigma=\Sigma(m,\sqrt{n})$, so that
evaluating $\partial_m \Sigma$ or $\partial_{\sqrt{n}} \Sigma$,
we only collect the log-divergent part contributing to the wave
function renormalization and this is
\beq
Z = 1- \frac1a \partial_{m} \Sigma|_{m=0}
 = 1 - \lambda S + O(\lambda^2)\,,\qquad  \qquad
S = \sum_{k}\frac{1}{(b\sqrt{k} +\mu)^2 }\,.
\eeq 
To $\Gamma_4$ contribute only a unique divergent four-point function
$F$ of the form given by Fig.\ref{fig:tad}. We have, still at one-loop, 
\beq
\Gamma_4(m,n,p,q) = - \lambda + \frac12\frac{\lambda^2}{2^2}
(2\cdot 2 \cdot 2)
\sum_{k} \frac{1}{(am + b\sqrt{k} + \mu)(ap+b\sqrt{k} + \mu)} + O(\lambda^3)\,.
\eeq
Note that, at this level, the above model
differs from the GW 4D model since more graphs
contribute now to the $\Gamma_4$ function. This entails a combinatorial factor twice greater \cite{Rivasseau:2007ab}.

We are in position to compute the $\beta$-function:
\beq
\lambda^{\text{ren}}
 = -\frac{\Gamma_4(0,0,0,0)}{Z^2}
 = -\frac{(-\lambda + \lambda^2S+ O(\lambda^3))}{(1- \lambda S+ O(\lambda^2))^2}
= \lambda + \lambda^2S + O(\lambda^3)\,.
\eeq
Therefore $\beta=+1$ and $\sigma T3$ is asymptotically
free in the UV as expected from \cite{BenGeloun:2012pu}.
In a similar way, all theories with propagator
$ap^{\frac12 + \epsilon} + bq^{\frac12}$, $\epsilon \in ]0,1/2]$
will be asymptotically free (once again, by the same reasons,
$T^-$ should be dropped because $\sum_{k}1/(ak^{\frac12 + \epsilon}+\mu)^2$ is convergent and the same contributions
of the four-point function are still convergent and should be neglected) whereas, at the end-point $\epsilon_0=0$, the $GW_{4D}$ model becomes safe.

\end{document}